\documentclass[aps,pre,twocolumn,showpacs,groupedaddress]{revtex4}

\usepackage{times}
\usepackage{graphicx}
\makeatletter
\newif\if@restonecol
\makeatother

\usepackage{algorithm2e}

\begin{document}

\title{A reconstruction algorithm for single-particle diffraction imaging experiments} 

\author{Ne-Te Duane Loh}
\author{Veit Elser}
\affiliation{Laboratory of Atomic and Solid State Physics Cornell University, Ithaca, NY 14853-2501} 

\date{\today}

\begin{abstract}
We introduce the EMC algorithm for reconstructing a particle's 3D diffraction intensity from very many photon shot-noise limited 2D measurements, when the particle orientation in each measurement is unknown. The algorithm combines a maximization step (M) of the intensity's likelihood function, with expansion (E) and compression (C) steps that map the 3D intensity model to a redundant tomographic representation and back again. After a few iterations of the EMC update rule, the reconstructed intensity is given to the difference-map algorithm for reconstruction of the particle contrast. We demonstrate reconstructions with simulated data and investigate the effects of particle complexity, number of measurements, and the number of photons per measurement. The relatively transparent scaling behavior of our algorithm provides a first estimate of the data processing resources required for future single-particle imaging experiments.
\end{abstract}


\maketitle


\section{Introduction}

If the goal of single-particle imaging by free electron x-ray lasers \cite{Neutze} is realized in the next few years, the disciplines of imaging and microscopy will have partly merged with elementary particle physics. Even with the enormous flux of the new light sources, the scattered radiation will be detected as individual photons and hardly resemble diffraction ``images" (Figure 1). The data in these experiments will instead resemble the particle debris produced in elementary particle collisions.

\begin{figure}[t]
\centering
\includegraphics[width=3.3in]{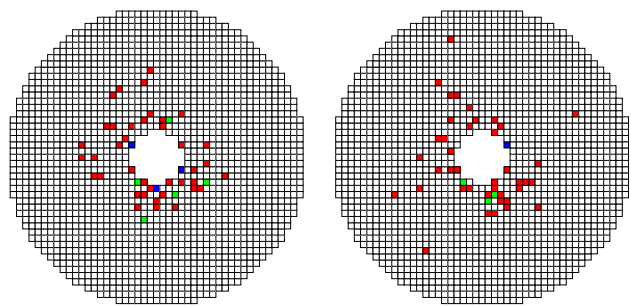} 
\caption{(Color online) The same or different? Two simulated measurements (noisy diffraction patterns) in a single particle imaging experiment, where color (white, red, green, blue) represents recorded photon counts (0, 1, 2, 3). Are the differences in the measurements purely statistical, or do they reflect a different view (orientation) of the particle? \label{fig:PhotonData}}
\end{figure}

The particle physics analogy is imperfect, however. Data analysis in elementary particle experiments is complicated more as the result of complex interactions than complexity of the structures --- consider the pions produced when a proton is probed with a photon. By contrast, the fundamental interactions between x-ray photons and electrons in a molecule are very simple and the complexity in the analysis of the data is entirely the result of structure.

There are two different data analysis challenges that x-ray laser studies of single-particles will have to face. Consider the two simulated detector outputs shown in Figure 1. Are the photon counts different because the molecule presented a different orientation to the x-ray beam; is the difference attributable to the statistics of a shot-noise limited signal; or does some combination of the two apply? It is reasonable to conjecture that by collecting sufficiently many data, the orientational and statistical uncertainties can be disentangled to produce molecular reconstructions with acceptable noise and resolution. In this paper we present strong evidence in support of this conjecture by means of an algorithm that succeeds with simulated data. 

Due to the length of this paper a survey of its contents may be useful to the reader. Section \ref{Theorysection} explains the theoretical basis of our algorithm whose success is contingent upon an information theoretic noise criterion. Sections \ref{Testparticlessection}, \ref{Expparamsection} and \ref{Reconparamsection} respectively describe the test target particles, experimental diffraction conditions and algorithmic parameters used in these single-particle imaging simulations. The limits and encouraging results of these simulations, whose code implementation we elaborate in section \ref{Algorithmsection}, are presented in section \ref{Simulationssection}. Finally, section \ref{Scalingsection} discusses the scaling of our algorithm's computational requirements with reconstructed resolution. Additional relevant technical details of our paper are recorded in its appendices.

\section{Theory}\label{Theorysection}

The statistical noise and missing orientational information can be addressed by imposing internal consistency of two kinds. First consider the shot noise of the detected signal. Suppose a collection of data sets, such as the pair in Figure 1, have been identified as candidates for data taken with the molecule in nearly the same orientation. While simply averaging the photon counts yields the continuous signal we are after, we have available the stronger test that the distribution of counts for the measurement ensemble, at each pixel, has the correct Poissonian form. If the test fails, then a different subset of the data must be identified which has this property. Statistical consistency, by itself, is thus a means for classifying like-oriented data sets.

The purely statistical analysis makes no reference to the structure implicit in the missing orientational information. This structure begins with the basic fact that the missing information comprises just three continuous variables (\textit{e.g.} Euler angles), and extends to more detailed constraints, such as the fact that the data samples a signal on a spherical (Ewald) surface in three dimensions and different spherical samples have common values along their intersection, \textit{etc}. A successful data analysis scheme for the single particle imaging experiments will not just have to signal-average shot noise, but must also reconstruct the missing orientational information by relying on internal consistency associated with the rotation group. The two forms of uncertainty, statistical and orientational, are not independent. In particular, when the statistical noise is large (few detected photons), we expect the reconstruction of the orientational information to be probabilistic in character (\textit{i.e.} distributions of angles as opposed to definite values).

\subsection{Noise criterion}\label{ssec:NoiseCriterion}

A natural question to ask is whether there exists an information theoretic criterion that would apply to any reconstruction algorithm and that can be used to evaluate the feasibility of reconstructions for particular experimental parameters. One of us \cite{Elser} studied this question for a minimal model with a single rotation angle and obtained an explicit noise criterion formula. Although such a detailed analysis is difficult to extend to the three dimensional geometry of the single-particle imaging experiments, the mathematical statement of the criterion is completely general and, when evaluated numerically by the reconstruction algorithm, serves as a useful diagnostic. We include a brief discussion of the criterion here and refer the reader to the original article for details.

We recall that the mutual information $I(X,Y)$ associated with a pair of random variables $X$ and $Y$ is an information theoretic measure of their degree of correlation: $I(X,Y)$ is the average information in bits that a measurement of $X$ reveals about $Y$ (or conversely). Keeping with the notation of references \cite{Elser,Huldt}, we denote the three dimensional intensity distribution by $W$, the photon counts recorded by the detector on a two dimensional spherical surface by $K$, and the three unknown parameters that specify the orientation of the surface within the intensity space by $\Omega$. The intensities $W$ have the interpretation of random variables (just as $K$ and $\Omega$), since the particle being reconstructed (and so the associated $W$) belongs to a statistical ensemble with known characteristics (size, intrinsic resolution, \textit{etc.}).

There are three forms of mutual information that arise in the framework where information about a model $W$ is obtained through measurement of data $K$ that is both statistically uncertain and incomplete (because $\Omega$ is not measured). The first is $I(K,W)$ and measures the information obtained about the model intensities $W$ from a typical unoriented measurement $K$. A second mutual information is $I(K,\Omega)|_W$, the correlation between the measurement $K$ and the orientation $\Omega$ conditional on a typical model $W$. We may also think of $I(K,\Omega)|_W$ as the entropy of $\Omega$ reduced by the finite entropy in its distribution when given typical measurements $K$ and models $W$. Finally, a simple identity (see appendix \ref{apd:mi}) yields the third form
\begin{equation}\label{threeMI}
I(K,W)|_\Omega = I(K,W)+I(K,\Omega)|_W
\end{equation}
as the sum of the other two. The mutual information $I(K,W)|_\Omega$ is the simplest of the three, as it measures the direct correlation between the continuous signal $W$ and its Poisson samples $K$ because it is conditional on a known orientation $\Omega$. In the limit where the mean photon count per detector pixel is much less than 1, this mutual information is given simply in terms of the total number of photons $N$ detected in an average measurement \cite{Elser},
\begin{equation}\label{MI1}
I(K,W)|_\Omega = (1-\gamma)N,
\end{equation}
where $\gamma\approx 0.577$ is Euler's constant.

In order to sufficiently sample the particle orientations and improve the signal-to-noise, information is accumulated over the course of many measurements. The information delivered in a stream of measurements will initially grow in proportion to the number of measurements, since typically each 2D measurement $K$ samples a different part of the 3D signal $W$. Two of the mutual information quantities introduced above may therefore be interpreted as \textit{information rates}:
\begin{eqnarray*}
I(K,W)|_\Omega&=&\mbox{data rate in a hypothetical experiment}\\
& &\mbox{with known particle orientations}\\
\\
I(K,W)&=&\mbox{data rate in the actual experiment}\\ 
&&\mbox{with unknown orientations}
\end{eqnarray*}
The time unit in these rates is the time for one measurement. The larger of these rates, $I(K,W)|_\Omega$, applies to the situation where the noisy data $K$ can simply be signal-averaged to obtain $W$. From the ratio
\begin{equation}
r=\frac{I(K,W)}{I(K,W)|_\Omega}
\end{equation}
we can assess the reduction in the data rate relative to the signal-averaging scenario. Because this reduction can be severe when shot noise is large, we are primarily interested in the dependence of $r$ on the mean photon number per measurement, $N$. Not only does an experiment with small $r(N)$ require a correspondingly larger number of measurements to obtain the same signal-to-noise in the reconstructed particle, our reconstruction algorithm (Section \ref{EMsection}) requires many more iterations in this case.

Upon using equation (\ref{threeMI}), the case $r(N)=1/2$ corresponds to the situation $I(K,W)=I(K,\Omega)|_W$, that is, the information in one unoriented measurement exactly matches the information acquired about its orientation. This interpretation does not imply that reconstruction is impossible for smaller $r(N)$, since the criterion refers to the properties of a single measurement while the reconstruction algorithm may, in principle, process many measurements in aggregate. Nevertheless, the criterion $r(N)>1/2$ correctly identifies the cross-over region separating easy and hard reconstructions. Using (\ref{threeMI}) we can rewrite the feasibility criterion in the form
\begin{equation}\label{criterion}
r(N)=1-\frac{I(K,\Omega)|_W}{I(K,W)|_\Omega}>\frac{1}{2}.
\end{equation}
Our algorithm, based on the expectation maximization principle \cite{EM}, evaluates $r(N)$ with no overhead since the probability distributions in $\Omega$ of the measurements $K$, from which $I(K,\Omega)|_W$ is derived, are computed in the course of updating the model. When the inequality above is strongly violated we should expect a much lower signal-to-noise in the finished reconstruction than a naive signal-averaging estimate would predict.

An important general observation about the noise criterion (\ref{criterion}) is that it is remarkably optimistic. As an information measure, $I(K,\Omega)|_W$ grows only logarithmically with the complexity of the particle. Recall that $I(K,\Omega)|_W$ is the entropy in $\Omega$ of a typical measurement $K$. Suppose a particle of radius $R$ has its density resolved to contrast elements of size $\delta R$. Its rotational structure (in its own space or the Fourier transform space of $W$) will then only extend to an angular resolution $\delta R/R$. A sampling of the rotation group comprising $(R/\delta R)^3$ elements thus provides a fair estimate of the entropy and $I(K,\Omega)|_W$ evaluates to a number of order $3\log{(R/\delta R)}$ \footnote{$I(K,\Omega)|_W$ also has a weak logarithmic variation with $N$ \cite{Elser}.}. This estimate and equations (\ref{MI1}) and (\ref{criterion}) imply that values of $N$ of only a few hundred should be sufficient to reconstruct even the most complex particles encountered in biology.

\subsection{Classification by cross correlating data}

The theoretical noise criterion discussed above is beyond the reach of the kind of algorithm that would seem to offer the most direct solution \cite{Huldt}. In this scenario the task is divided into two steps. The first is concerned with classifying the diffraction data into sets that, with some level of confidence, describe the particle in a  small range of orientations. After averaging photon counts for the like-sets to improve the signal quality, the diffraction pattern averages would then be assembled into a consistent three dimensional intensity distribution in the second step.

The most direct method of assessing the similarity of two diffraction data is to compute the cross correlation. A pair of like-views of the particle would be identified by a large cross correlation. Because this measure also fluctuates as the result of shot noise, its statistical significance must be estimated as well. The result of such an analysis \cite{Huldt} is that the cross correlation based classification can succeed only if the average number of photons per diffraction pattern, $N$, and the number of detector pixels, $M_\mathrm{pix}$, satisfy
\begin{equation}
N\gg\sqrt{M_\mathrm{pix}}.
\end{equation}
This criterion imposes a higher threshold on $N$ than the information theoretic criterion (\ref{criterion}) because $M_\mathrm{pix}$ grows algebraically, and not logarithmically, with the complexity of the reconstructed particle.

Since the number of measured diffraction patterns will be very large, and the number of pairs to be cross correlated grows as its square, the execution of this approach also seems prohibitive. As Bortel \textit{et al.} \cite{Bortel} have shown, however, this estimate is overly pessimistic since by selecting suitable representatives of the orientational classes the number of cross correlation computations can be drastically reduced. The expectation maximization (EM) algorithm described below is an alternative classification method where the most time consuming step is again the computation of very many cross correlations. But unlike methods where both vectors of the cross correlation are data and criterion (5) applies, in the EM algorithm only one of the vectors is data while the other is derived from a model. This has the added bonus that the time of the EM calculation is linear, rather than quadratic, in the number of measurements.

\subsection{Classification by expectation maximization}\label{EMsection}

The algorithm we have developed for the single particle imaging experiments and studied previously in the context of noise limits \cite{Elser} is based on the idea of expectation maximization (EM) \cite{EM} . In general, EM seeks to reconstruct a model from statistical data that is incomplete. The model in the present setting is the intensity signal $W$, the data are the sets of photon counts $K$ recorded by the detector, and the latter are incomplete because the orientation $\Omega$, of the particle relative to the detector, is not measured.

The EM algorithm is an update rule on the model, $W\to W^\prime$, based on maximizing a log-likelihood function $Q(W^\prime)$. The algorithm derives its name from the fact that $Q(W^\prime)$ is actually an expectation value of log-likelihood functions, where a probability distribution based on the current model parameters $W$ is applied to the missing data $\Omega$. We will derive $Q(W^\prime)$ for the single particle imaging problem in stages, beginning with the log-likelihood function for the photon counts at a single detector pixel.

Let $W(\mathbf{q})$ be the time-integrated scattered intensity at spatial frequency $\mathbf{q}$ when the particle is in some reference orientation. The detector pixels, labeled by the index $i$, approximately measure $M_\mathrm{pix}$ point samples $W(\mathbf{q}_i)$. When multiplied by the pixel area and divided by the photon energy, $W(\mathbf{q}_i)$ corresponds to the average photon number recorded at pixel $i$. Since these normalization factors are constants, we will refer to $W$ interchangeably as ``intensity" or ``average photon number."
If we now give the particle some arbitrary orientation $\Omega$, the average photon number at pixel $i$ is $W(\mathbf{R}_\Omega\cdot\mathbf{q}_i)$, where $\mathbf{R}_\Omega$ is the orthogonal matrix corresponding to the rotation between the reference orientation and $\Omega$. Because the implementation of the algorithm approximates the continuous $\Omega$ with a discrete sampling of $M_\mathrm{rot}$ points labeled by the index $j$, we define $W_{ij} = W(\mathbf{R}_j\cdot\mathbf{q}_i)$ as the average photon number at detector pixel $i$ when the particle has orientation $j$.

The log-likelihood function for the mean photon number $W'_{ij}$, given a photon count $K_{ik}$ at pixel $i$ in measurement $k$, is the logarithm of the Poisson distribution (apart from an irrelevant constant):
\begin{equation}
Q_{ijk}(W^\prime)= K_{ik} \log{W_{ij}^\prime}-W_{ij}^\prime.
\end{equation}
Summing this function over the detector pixels gives the log-likelihood function associated with the joint and independent Poisson distributions on the photons detected in a single measurement (labeled by $k$):
\begin{equation}
Q_{jk}(W^\prime)= \sum_{i=1}^{M_\mathrm{pix}} Q_{ijk}(W^\prime).
\end{equation}
If we knew the orientation $j$ that applied to the counts $K_{ik}$ of measurement $k$, we would try to maximize the corresponding $Q_{jk}$ with respect to the model values $W_{ij}^\prime$. The EM algorithm deals with the missing information by making an educated estimate of $j$, for each measurement $k$, based on the current model values. However, before we enter into these details we should point out that the EM algorithm in our formulation works with many more model parameters than there are in the physical model. That is because $W_{ij}$ and $W_{i^\prime j^\prime}$ are treated as independent parameters even in the event that the corresponding spatial frequencies $\mathbf{R}_j\cdot\mathbf{q}_i$ and $\mathbf{R}_{j^\prime}\cdot\mathbf{q}_{i^\prime}$ are nearly the same. This overspecification of parameters will be rectified by the ``compression step" described below.

The EM algorithm defines the log-likelihood function $Q(W^\prime)$ on the updated model parameters $W^\prime$ by assigning a provisional distribution of orientations $j$ to each measurement $k$ based on the current model parameters $W$. The $j$-distribution is given as the normalized likelihood function for the measurements $K_{ik}$ conditional on $j$ and the model parameters $W$. Up to an irrelevant $j$-independent factor, the conditional probability in question is the product of Poisson probabilities at each detector pixel:
\begin{equation}\label{Rjk}
R_{jk}(W)=\prod_{i=1}^{M_\mathrm{pix}} W_{ij}^{K_{ik}}\exp{(-W_{ij})}.
\end{equation}
The normalized likelihood function allows for an arbitrary prior distribution of the orientations $j$ which we denote by the normalized weights $w_j$:
\begin{equation}
P_{jk}(W) = \frac{w_j R_{jk}(W)}{\sum_j w_j R_{jk}(W)}.
\end{equation}
This form is necessary even when the prior distribution on orientations $\Omega$ is uniform (the usual assumption for the single particle experiments) because in general the discrete samples $j$ cannot be chosen in such a way that the weights $w_j$ are uniform. The EM log-likelihood function may now be written explicitly:
\begin{equation}\label{eqn:likelihood}
Q(W^\prime)=\sum_{k=1}^{M_\mathrm{data}}\sum_{j=1}^{M_\mathrm{rot}}P_{jk}(W) Q_{jk}(W^\prime).
\end{equation}

Details on the maximization of $Q(W^\prime)$ are given in appendix \ref{apd:EM}; the resulting ``maximizing" ($\mathbf{M}$) update rule is simple and intuitive:
\begin{equation}\label{Mrule}
\mathbf{M}:\quad W_{ij}\to W_{ij}^\prime =\frac{\sum_{k=1}^{M_\mathrm{data}}P_{jk}(W) K_{ik}}{\sum_{k=1}^{M_\mathrm{data}}P_{jk}(W)}.
\end{equation}
We see that the data $K_{ik}$ are averaged over all the data sets ($k$ index) with the unknown orientation index $j$ distributed according to probabilities $P_{jk}(W)$ defined by the current model.
It is instructive to check that the update rule applied to an arbitrary rotation of the true signal leaves the signal unchanged (details in appendix \ref{apd:EM}). 

We now return to the point that the parameters $W_{ij}$ overspecify the true model parameters. For fixed $j$, the $M_\mathrm{pix}$ numbers $W_{ij}$ correspond to a tomographic sampling of the 3D space of intensities on a spherical surface with orientation in the 3D space specified by $j$. To recover a signal in the 3D space we define a ``condensation/compression" ($\mathbf{C}$) mapping
\begin{equation}
\mathbf{C}:\quad W_{ij}\to  W(\mathbf{p}), 
\end{equation}
where $\mathbf{p}$ denotes a spatial frequency sampling point in the 3D intensity space. Since the samples $\mathbf{p}$ will be arranged on a regular 3D grid, we define interpolation weights $f(\mathbf{q})$ for a general point $\mathbf{q}$ in the 3D space which vanish for large $|\mathbf{q}|$ and have the property
\begin{equation}\label{eqn:interp}
1=\sum_\mathbf{p} f(\mathbf{p}-\mathbf{q})
\end{equation}
for arbitrary $\mathbf{q}$. Recalling that the value $W_{ij}$ corresponds to the 3D sampling point $\mathbf{R}_j\cdot\mathbf{q}_i$, the signal values after the compression mapping are given by
\begin{equation}\label{compress}
W(\mathbf{p})=\frac{\sum_{i=1}^{M_\mathrm{pix}}\sum_{j=1}^{M_\mathrm{rot}} f(\mathbf{p}-\mathbf{R}_j\cdot\mathbf{q}_i) W_{ij}}{\sum_{i=1}^{M_\mathrm{pix}}\sum_{j=1}^{M_\mathrm{rot}} f(\mathbf{p}-\mathbf{R}_j\cdot\mathbf{q}_i)}.
\end{equation}

To begin another round of the EM algorithm, after the condensation step, the signal values on the 3D grid have to be ``exported/expanded" ($\mathbf{E}$) to the tomographic representation:
\begin{equation}
\mathbf{E}:\quad W(\mathbf{p})\to W_{ij}^\prime.
\end{equation}
Using the same interpolation weights and rotation samples $j$ as before, we have
\begin{equation}
W_{ij}^\prime= \sum_\mathbf{p}f(\mathbf{p}-\mathbf{R}_j\cdot\mathbf{q}_i) W(\mathbf{p}).
\end{equation}
The combined mappings $\mathbf{E}\cdot \mathbf{C}: W_{ij}\to W_{ij}^\prime$ then have the effect of imposing on the redundant tomographic representation of the signal the property that it is derived from values on a 3D intensity grid. A slightly different way of grouping the three mappings defines one iteration of what we will call the EMC algorithm (\textbf{E}xpansion followed by expectation \textbf{M}aximization followed by \textbf{C}ompression):
\begin{equation}
\mathbf{C}\cdot \mathbf{M}\cdot \mathbf{E}:\quad W(\mathbf{p})\to W^\prime(\mathbf{p}).
\end{equation}

The most time consuming step of the EMC algorithm is the computation of the probabilities $P_{jk}(W)$. Prior to normalization these are the likelihood functions $R_{jk}(W)$ whose logarithms are given by
\begin{equation}
\log{\left(R_{jk}(W)\right)}=\sum_{i=1}^{M_\mathrm{pix}}K_{ik} \log{W_{ij}}-W_{ij}.
\end{equation}
At the heart of the algorithm we have to compute the cross correlation (sum on $i$) between the photon counts in each measurement $k$ and the logarithm of the signal at each tomographic sampling (particle orientation) $j$. Since data are not cross correlated with data, as in some classification methods, the time scaling is linear in the number of measurements. After normalizing to get  $P_{jk}(W)$, the mutual information needed for the noise criterion (\ref{criterion}) is obtained without significant additional effort (details in appendix \ref{apd:EM}):
\begin{equation}\label{eqn:MIKOmega}
I(K,\Omega)|_W = 
\frac{1}{M_\mathrm{data}}\sum_{k=1}^{M_\mathrm{data}}\sum_{j=1}^{M_\mathrm{rot}}P_{jk}(W)\log{\left(\frac{P_{jk}(W)}{w_j}\right) }.
\end{equation}

The expectation maximization technique described above is very similar to that used by Scheres \textit{et al.} \cite{Scheres}  for cryo-EM reconstructions. Cryo-EM and single-particle x-ray imaging differ in two important physical respects. The first is that the diffraction data in the x-ray experiments has a known origin (zero frequency), thereby reducing the missing information. This is not completely an advantage because the diffraction data, after a successful reconstruction, must undergo an additional stage of phase retrieval before the results can be compared with cryo-EM. The second difference is the noise model that applies to the two techniques. In the absence of background, the shot noise in the \mbox{x-ray} experiments is a fundamental and parameter-free process, whereas the background ice scattering in cryo-EM requires phenomenological models. The expectation maximization algorithm is general enough that these differences do not change the overall structure of the reconstruction process. In fact, the work of Scheres \textit{et al.}  \cite{Scheres} points out that the algorithm is readily adapted to include additional missing data, such as conformational variants of the molecule.

Redundant representations of the model parameters, and operations that impose consistency with a physical (3D) model, are also shared features. Scheres \textit{et al.} \cite{Scheres} obtain the 3D model using ART \cite{ART}, a least-squares projection technique. The corresponding operation in our reconstruction algorithm is the linear compression-expansion mapping $\mathbf{E}\cdot \mathbf{C}$. The redundancy question did not arise in the same way for the minimal model studied previously \cite{Elser}, with only a single rotation axis. There the intensity tomographs did not intersect and the speckle structure had to be imposed by an additional support constraint on the Fourier transform of the intensity distribution.

Finally, we note that Fung \textit{et al.} \cite{Ourmazd} have developed a technique that, like our expectation maximization approach, uses the entire body of data in a single update of the model parameters.

\section{Test particles}\label{Testparticlessection}

When simulating the single-particle experiments it is important to distinguish between the resolution limit imposed by the maximum measured scattering angle and the resolution intrinsic to the scattering particle as a result of its dynamics. The scattering cross section for a complex molecule is generally significantly smaller, at large momentum transfer, than what is predicted by atomic form factors and a static molecular structure. This phenomenon is well known in crystallography, where the coherent illumination of numerous molecules in various states of perturbation is equivalent --- when considering the information recorded in Bragg peaks --- to a single molecule with blurred contrast. The same effect, but in the temporal domain, will diminish the scattering at large angles in the single-particle experiments.

The dynamics of molecules subject to intense x-ray pulses is complicated by the presence of several physical processes \cite{explosion}. In the case where the degree of ionization of the atoms is relatively low during the passage of the pulse, the x-ray scattering is dominated by the atomic cores and can be analyzed by modeling the atomic motions. Let $A_\mathbf{k}(t)$ be the amplitude of the incident radiation in a particular momentum mode $\mathbf{k}$. The cross section for scattering a photon into mode $\mathbf{k}+\mathbf{q}$ with frequency $\omega$ contains as a factor the expression
\begin{equation}\label{molecule_form_factor}
\Gamma(\mathbf{q}) = \left| \int dt\, A_\mathbf{k}(t)e^{i \omega t}\sum_p f_p(\mathbf{q})\exp{\left[i \mathbf{q}\cdot \mathbf{r}_p(t)\right]}\right|^2,
\end{equation}
where $f_p(\mathbf{q})$ is the atomic form factor of atom $p$ whose position $\mathbf{r}_p(t)$ changes with time as a result of large scale ionization, \textit{etc}. We are interested in modeling the $\mathbf{q}$-dependence of the molecular form factor (\ref{molecule_form_factor}). Without access to detailed simulations of the Coulomb explosion, we have adopted for our data modeling the simple one-parameter form,
\begin{equation}\label{Gauss_form_factor}
\Gamma(\mathbf{q})/\Gamma(0) = \exp{(-B |\mathbf{q}|^2)}\, S(\mathbf{q}),
\end{equation}
where $S(\mathbf{q})$ is the normalized static ($t=0$) structure factor of the molecule.
This Gaussian form results if the dynamics of the positions $\mathbf{r}_p(t)$ can be approximated by independent Gaussian fluctuations over the coherent time scale $T$ of the pulse. The period $T$, or the time during which the function $A_\mathbf{k}(t)e^{i \omega t}$ is approximately constant, is significantly shorter than the pulse duration in a non-seeded free electron laser \cite{non-seeded}.

We expect more detailed dynamical simulations of the scattering cross section to show significant departures from the form above, of a simple Gaussian factor modulating a static structure factor. Rather, the effective structure factor of an exploding molecule should resemble that of an atomistic density that is primarily blurred radially, with contrast at the surface of the molecule experiencing the greatest degradation \cite{explosion}. Given such complications, our test particle modeling ignores atomicity and treats the particle more simply as a distribution of positive contrast on a specified support with a phenomenologically defined intrinsic resolution given by the form (\ref{Gauss_form_factor}).

\subsection{Binary contrast particles}

It is a great convenience, when developing algorithms, to have a simply defined ensemble of problem instances that offers direct control over the key parameters. We have chosen to work with an ensemble having a single parameter that controls the complexity of the particle, where our measure of complexity is the dimensionless radius $R$ which specifies the physical particle radius in units of the intrinsic resolution. Our particles have the following properties:
\begin{center}
\parbox{3in}{(1) Spherical shape, (2) binary contrast, and (3) Gaussian form factor.}
\end{center}
Property (1) is chosen to make the reconstructions as hard as possible, both for the assembly of the 3D intensity and later in the phase retrieval stage (the spherical support offering the fewest constraints \cite{ElserMillane}). We chose (2) to mimic a large biomolecule that because of damage can only be resolved into solvent (empty) and non-solvent regions. This property also has the convenience that most of the information about the structure is conveyed by rendering a single 3D contour at an intermediate contrast. Property (3) defines the intrinsic resolution.

The construction of a test particle begins by choosing a value for the dimensionless radius $R$; the final contrast values will be defined on a cubic grid of size $2R+1$. Our test particle construction algorithm is diagramed in pseudocode (algorithms \ref{alg:particle_constr} and \ref{alg:binary_contrast}) . It uses the particle support $S$ (voxels inside the sphere of radius $R$) and a Gaussian low-pass filter $F(\mathbf{q})$ to impose the form factor (\ref{Gauss_form_factor}) on the Fourier transform of the contrast. To keep the dynamic range of the intensity measurements in our simulations constant for different particle sizes $R$, we parameterize the filter as 
\begin{equation}\label{filter}
F(\mathbf{q})=\exp{\left[-1.5 (|\mathbf{q}|/q_\mathrm{max})^2\right]},
\end{equation}
where $q_\mathrm{max}=\pi/R$. Since only scattering with $|\mathbf{q}|<q_\mathrm{max}$ is measured, the discarded power is always
\begin{equation}
\frac{\int_1^\infty \exp{(-3 q^2)}\, q^2 dq}{\int_0^\infty \exp{(-3 q^2)}\, q^2 dq}=11\%.
\end{equation}

\begin{figure}[t!]
\centering
\includegraphics[width=3.3in]{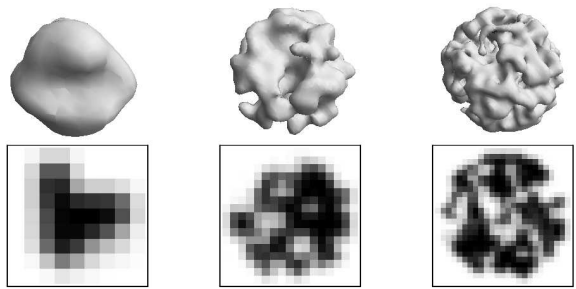}
\caption{Examples of the random binary contrast particles used in our simulations. The contrast is nearly uniform inside a labyrinthine region that fills half the volume of a sphere. Shown are particles of radius $R=4,8,12$ (left to right), where the dimensionless $R$ is in units of the intrinsic resolution, \textit{i.e.} the scale of the voxels shown in cross-section for each particle in the lower panel. The particle renderings in this paper, such as those above, show the iso-contrast surface appropriate for the labyrinth walls (half the maximum contrast). \label{fig:testparticles}}
\end{figure}

\restylealgo{boxed}

\begin{algorithm}[t] 
\SetLine
\SetInd{0.5em}{0.5em}
\caption{test particle construction \label{alg:particle_constr}}

\BlankLine
\KwOut{particle contrast on cubic grid $C$}				

\SetKwFunction{RandomContrast}{RandomContrast}
\SetKwFunction{BinaryContrast}{BinaryContrast}
\SetKwFunction{LowPassFilter}{LowPassFilter}

\BlankLine

$C\leftarrow \RandomContrast$

\BlankLine

\For{$i\leftarrow 1$ \KwTo $4$}
	{
	$C\leftarrow\BinaryContrast{C}$

	$C\leftarrow\LowPassFilter{C}$
	}

\BlankLine
\Return $C$
\BlankLine
\end{algorithm}

\begin{algorithm}[h] 
\SetLine
\SetInd{0.5em}{0.5em}
\SetAlgoSkip{smallskip}
\caption{binary contrast projection \label{alg:binary_contrast}}

\SetKwFunction{BinaryContrast}{BinaryContrast}
\SetKwFunction{MedianValue}{MedianValue}

\BlankLine
\KwIn{arbitrary contrast on $C$, support $S$}		
\KwOut{$\BinaryContrast (C)$}

\BlankLine

\lForEach{$\mathbf{r}\notin S$}{ 
		$C[\mathbf{r}]\leftarrow 0$} 
		
\BlankLine
		
$v\leftarrow \MedianValue({C[\mathbf{r} \in S]})$

\BlankLine

\ForEach{$\mathbf{r}\in S$}{ 
		\eIf{$C[\mathbf{r}]<v$}
		{$C[\mathbf{r}]\leftarrow 0$ }{ $C[\mathbf{r}]\leftarrow 1$ }
		}
\BlankLine
\Return $C$
\BlankLine
\end{algorithm}

In Figure 2 we show examples of binary contrast test particles constructed for three values of the dimensionless radius $R$. The largest particles considered in this study had $R=8$ because the reconstruction computations grow, in both time and memory, very rapidly with $R$. In Section 7 we discuss the scaling of the computations with $R$.

\subsection{Degraded resolution biomolecules}

To put our dimensionless radius $R$ in perspective, the same low-pass filter used for test particles was applied to the roughly $0.8$ MDa biomolecule GroEL \cite{GroEL}. After binning the coordinates of the non-hydrogen atoms of the PDB structure on a cubic grid of resolution 2\AA, the discrete Fourier transform was computed and truncated at the size $2R+1$. The result was then multiplied by the filter (\ref{filter}) and inverse transformed to give the contrast used in the simulation.

Figure 3 shows GroEL processed in this way for three values of $R$. Handedness in the protein secondary structure begins to appear at about $R=6$. These degraded resolution models of GroEL, that mimic the effects of dynamics and a finite duration pulse, are of course completely phenomenological. It may not even be true that the diffraction signal can be modeled by an appropriately blurred contrast function. This will be the case, for example, if the damage dynamics strongly varies with the orientation of the particle. Finally, we cannot ignore the fact that, as a result of thermal motion and solvent, at some level of resolution even the model of a unique ($t=0$) diffraction signal breaks down.

\begin{figure}[t!]
\centering
\includegraphics[width=3.3in]{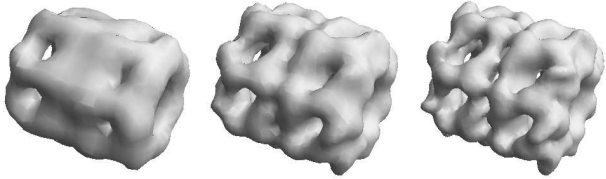}
\caption{Contrast of the protein complex GroEL \cite{GroEL} degraded by the same type of low-pass filter used in the construction of random test particles (Fig. \ref{fig:testparticles}). The axial length of GroEL is approximately 15 nm. At high damage, $R=4$ (left), the contrast reveals only the gross particle shape (``cornEL"). Handedness of the protein secondary structure appears at about $R=6$ (middle) and is fully evident by $R=8$ (right). \label{fig:blurredGroEL}}
\end{figure}

\section{Experimental parameters}\label{Expparamsection}

\subsection{Detector parameters}

The detector geometry, pixel dimensions, and position relative to the scattering particle determine the spatial frequency samples $\mathbf{q}_i$ of the experimental data. Our simulations use a detector model with three parameters: oversampling factor $\sigma$, maximum scattering angle $\theta$, and a dimensionless central data cutoff $\alpha$.

The oversampling factor has the most direct interpretation in real space. Oversampling $
\sigma$ corresponds to embedding the particle, with contrast defined on a grid of size $2R+1$, on a grid magnified in size by the factor $\sigma$. This also defines the dimensions of the 3D intensity grid, on which most of the computations of the EMC algorithm take place. Since Fourier transforms play no role in these computations there is no incentive to make the dimensions of the intensity grid a product of small primes. It is more natural in the EMC calculations, which have rotational symmetry about zero frequency, to have intensity grids of odd dimension with indices that run between $-q_\mathrm{max}$ and $+q_\mathrm{max}$. Here $q_\mathrm{max}$ is given by $\sigma \,R$ rounded up to the nearest integer. Speckles in the intensity distribution will have a linear size $\sigma$ in grid units.

A real detector does not measures point samples with respect to spatial frequency but convolves the true intensity signal with a point spread function defined by the pixel response \cite{detectorPSF}. To minimize this effect the oversampling in experiments should be kept large. Another reason for keeping $\sigma$ large is algorithmic: the expansion and compression steps of the EMC algorithm, which interpolate between tomographic and grid samples, introduce errors that are also minimized when the oversampling is large. In this study we used $\sigma=6$.

The maximum scattering angle is determined by the radius of the detector, $L$, and the distance $D$ between the particle and the detector, by $\tan{\theta}=L/D$. We define $L$ to be the radius of the largest disk that fits inside the actual detector. This corresponds to discarding data recorded in pixels outside the disk, in the corners of the detector. With this minor truncation of the data, all of it can be embedded in the 3D intensity grid for any particle orientation (relative to the reference orientation).

The actual choice of spatial frequencies $\mathbf{q}_i$ used by the EMC algorithm is largely arbitrary, and so we start by considering detector pixels at arbitrary positions $[x_i, y_i]$ in the detector plane. Up to a constant factor, the photon momentum detected at pixel $i$ is
\begin{equation}\label{momentum}
\mathbf{p}_i=\frac{[x_i,\,y_i,\,D]}{\sqrt{x_i^2+y_i^2+D^2}},
\end{equation}
and the corresponding spatial frequency, or momentum transfer from the incident beam, is $\mathbf{q}_i=\mathbf{p}_i-\mathbf{p}_0$, where $\mathbf{p}_0$ is given by (\ref{momentum}) with $x_i=y_i=0$. Intensities at these spatial frequencies will be represented as interpolated values with respect to the 3D intensity grid. Since the latter has unit grid spacing, we choose an appropriate rescaling of the $\mathbf{q}_i$ that is well matched with this. Because most detectors will have pixel positions on a square lattice with some pixel spacing $d$, our simulations are based on this model. We note, however, that the EMC algorithm operates with general tables of frequencies $\mathbf{q}_i$, and whether these are derived from a standard detector or a more complex tiled design is invisible to the workings of the algorithm. For the square-array detector $x_i= m_i\, d$, $y_i=n_i\, d$, where $m_i$ and $n_i$ are integers, and the rescaled spatial frequencies are given by
\begin{equation}\label{qi}
\mathbf{q}_i=\frac{[m_i,\,n_i,\,D/d]}{\sqrt{(m_i^2+n_i^2)(d/D)^2+1}}-[0,0,\,D/d].
\end{equation}
These samples lie on the surface of a sphere that passes through the origin and the scaling is such that samples near the origin match the 3D integer grid:
\begin{equation}
\mathbf{q}_i\approx [m_i,\,n_i,\,0].
\end{equation}
The pixels at the edge of the detector have
\begin{equation}
\sqrt{m_i^2+n_i^2}=L/d,
\end{equation}
and should correspond to frequencies at the highest resolution shell, or $|\mathbf{q}_i|=q_\mathrm{max}$. This condition, evaluated for (\ref{qi}) with $D=L \cot{\theta}$, reduces to
\begin{equation}\label{Ewald}
q_\mathrm{max}=(L/d)\cos{\theta}\sec{(\theta/2)}.
\end{equation}
For a small maximum scattering angle this reduces to the equality between the pixel size of the detector, $2(L/d)$, and the number of samples in one dimension of the intensity grid, $2 q_\mathrm{max}$. The $\theta$-dependence of expression (\ref{Ewald}) is a result of the spherical shape of the Ewald sphere.

\begin{figure}[t]
\centering
\includegraphics[width=3.3in]{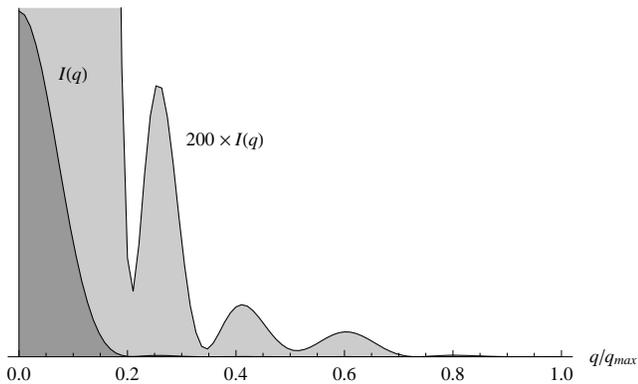} 
\caption{Radial intensity scan for an $R=8$ test particle on a linear scale. Our simulations only use data collected outside the central speckle, $q/q_\mathrm{max}>\alpha/R=0.18$.\label{fig:centspeck}}
\end{figure}

The forward scattering by the uniform or uninteresting part of the charge density of a compact particle is so much more intense than the scattering at larger angles from the non-uniform, interesting part, that most detectors need to have the central pixels blocked (beyond what is needed to avoid the incident beam). Figure \ref{fig:centspeck} shows a simulated intensity scan passing through the origin, for one of the test particles described in Section \ref{Testparticlessection}. The huge central speckle contains essentially only information about the total charge and almost no information about the structure of the particle. A natural size for the detector block is such that scattering at frequencies inside the main speckle is discarded. We obtain the cutoff frequency $q_\mathrm{min}$ by evaluating the scattering amplitude of a uniform ball having the same radius $R$ as the test particle. The first vanishing of this amplitude determines $q_\mathrm{min}$:
\begin{equation}
 (R/q_\mathrm{max})q_\mathrm{min}=\alpha
\end{equation}
where $\alpha\approx 1.43$ is the first non-zero root of $\pi x=\tan{\pi x}$. Since $q_\mathrm{max}=\sigma \,R$, more generally we define
\begin{equation}
q_\mathrm{min}=\sigma\,\alpha,
\end{equation}
which shows that the low frequency cutoff is $\alpha$ times the speckle size in grid units ($\sigma$).

We conclude this section by reviewing the procedure for generating the spatial frequencies $\mathbf{q}_i$ used by the EMC algorithm. Prior to this a dimensionless test particle radius $R$ has been selected. The half-size of the intensity grid is then given by $q_\mathrm{max}=\sigma\, R$, and our simulations used $\sigma=6$. Given the maximum scattering angle $\theta$ we then determine the detector radius in pixel units, $L/d$, from (\ref{Ewald}), as well as the detector-particle distance $D/d$ using $D=L\cot{\theta}$. All our simulations used $\theta=45^\circ$. Having determined $L/d$, we determine (for our choice of square array detector) the indices $m_i$, $n_i$ satisfying $m_i^2+n_i^2<(L/d)^2$. These are used in formula (\ref{qi}) to give the table of frequency samples in the reference orientation of the particle. Finally, to model the discarded central data we remove from the table all samples with $|\mathbf{q}_i|<q_\mathrm{min}$, where $q_\mathrm{min}= 1.43\sigma$.

\subsection{Diffracted signal strength}\label{ssec:signalstrength}
A key experimental parameter is the flux of photons incident on the particle. For the purpose of simulating the reconstruction process, however, a more convenient form of this parameter is the average number of photons scattered to the detector in one measurement, $N$. This normalization of the diffraction signal can be carried out once the detector's spatial frequency samples $\mathbf{q}_i$ are determined.

In order to generate diffraction data with the property that the mean photon number is $N$, we first compute the squared magnitude of the Fourier transform of the particle contrast embedded on the intensity grid (having size $2 q_\mathrm{max}$). We interpret the numbers on this grid as the photon flux scattered into the respective spatial frequencies, at this point with arbitrary global normalization. A detector pixel at one such frequency sample will record an integer photon count drawn from the Poisson distribution having the (time and pixel area-integrated) flux as mean. The quantity we wish to normalize is the net flux at all the detector pixels. When this quantity is $N$, then the mean photon number per measurement will also be $N$.

Because the particle contrast is not spherically symmetric, the net diffracted flux to the detector pixels will fluctuate with the particle orientation. We avoid bias arising from this effect by sampling a few hundred orientations of the particle and applying the associated rotations to the frequency samples in order to estimate the orientation-averaged flux. This number is then used to rescale the flux values on our 3D intensity grid. With this normalization in place, we generate data by repeatedly sampling random orientations, rotating the frequency samples, and then drawing Poisson samples at each of the rotated frequencies for the means given on the normalized intensity grid. Linear interpolation of the grid intensities is used to obtain the diffracted fluxes at the rotated frequencies.

\section{Reconstruction parameters}\label{Reconparamsection}

In addition to the diffraction data, prior information about the particle provides additional parameters to the reconstruction algorithm. The only such information we consider in our simulations is the dimensionless particle radius $R$. This parameter is used in both the intensity reconstruction by the EMC algorithm, as well as the phase retrieval stage that reconstructs the particle contrast from the intensity. 

\subsection{Rotation group sampling}\label{Rotationsection}

The EMC algorithm orients 2D data tomographs within the 3D intensity distribution using a discrete sampling of the rotation group. An optimal sampling is one where the samples are uniformly distributed and at a sufficient density to resolve the smallest angular features. Because speckles in the intensity distribution have linear dimension $\sigma$, features of this size (in voxel units) at the highest resolution shell, $q_\mathrm{max}$, determine the angular scale:
\begin{equation}\label{angularRes}
\delta\theta=\sigma/q_\mathrm{max}=1/R.
\end{equation}

\begin{figure}[t]
\centering
\includegraphics[width=3.3in]{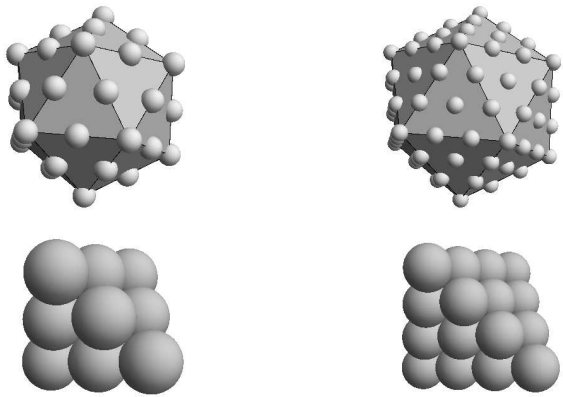} 
\caption{Sampling the 3D rotation group is equivalent to sampling the surface of a sphere in 4D. Shown in the top row is a scheme for sampling the surface of a sphere in one lower dimension based on subdivisions of the 20 faces of the icosahedron. The analogous construction in one higher dimension subdivides the 600 tetrahedral faces of the 600-cell, two examples of which are shown in the bottom row. The resolution of the sampling is specified by the number of subdivisions of each edge; shown are $n=2$ (left) and $n=3$ (right). \label{fig:cannonballs}}
\end{figure}

The rotation group parameterization that is best suited for generating uniform samplings is based on quaternions \cite{quaternions}. Unit quaternions are points on the unit sphere in four dimensions and encode 2-to-1 the elements of the continuous rotation group in three dimensions. Their key property is the fact that the distance between quaternions $q$ and $q^\prime$, in the usual sense, is simply related to the angle of the relative rotation between the group elements associated with $q$ and $q^\prime$. For small relative rotations $\delta\theta$ this relationship is:
\begin{equation}
 \|q-q^\prime\|\approx \delta\theta/2.
\end{equation}
Given a $\delta\theta$, the problem of selecting rotation group samples, such that any rotation is within a relative rotation $\delta\theta$ of some sample, is thus equivalent to the standard problem of constructing efficient coverings \cite{coverings} of the 3-sphere. We solve this covering problem by using a design based on a highly symmetric polytope, the 600-cell \cite{Coxeter}. This polytope is the four dimensional analog of the icosahedron in that it approximates the curved surface of the sphere by a union of regular simplices --- 3D tetrahedra rather than triangles in four dimensions. The regular tetrahedron is efficiently covered by points in the fcc arrangement; coverings with increasing resolution are shown in Fig. \ref{fig:cannonballs}. The resolution of the covering is parametrized by an integer $n$ that gives the number of subdivisions of each edge of the tetrahedron. Our 3-sphere coverings are obtained by rescaling the points that cover the tetrahedral faces of the 600-cell to unit length. Details of the construction, including the computation of the sample weights, are given in appendix \ref{apd:rot}.

There are only two properties of the rotation group sampling that have direct relevance to the EMC algorithm for intensity reconstruction: the angular resolution $\delta\theta$ and the number of rotation samples, $M_\mathrm{rot}$. Defining $\delta\theta$ as the covering radius of the sampling, the $n$-dependence is given by (see appendix \ref{apd:rot}):
\begin{equation}
\delta\theta(n)\approx 0.944/n.
\end{equation}
When combined with the estimate (\ref{angularRes}), this implies that $n$ should roughly coincide with the dimensionless particle radius $R$. Moreover, since $n\approx R$, the number of samples (see appendix \ref{apd:rot}),
\begin{equation}\label{eqn:Mrot}
M_\mathrm{rot}(n)=10(5 n^3+n),
\end{equation}
grows in proportion with the volume of the particle. 

\subsection{Particle support}\label{ssec:support}

Our phase reconstruction of the complex diffraction amplitude is carried out with the diffraction magnitude on the same grid as used by the EMC algorithm for the intensity reconstruction. The support constraint is therefore that the particle contrast can be non-zero only within a sphere of radius $R$ grid units. In our simulations, which were limited to $R\le 8$, we increased the support radius by one or two units because precise knowledge of the support is usually not available in real experiments.

\section{Reconstruction algorithm}\label{Algorithmsection}

Our algorithm for reconstructing the scattering contrast of a particle begins by reconstructing the 3D intensity with the EMC algorithm for classifying diffraction data. This section describes in concise algorithmic language the EMC process already sketched in Section \ref{EMsection}. For the relatively much easier final step, of reconstructing the particle contrast from the intensity, we use the difference-map (DM) phase reconstruction algorithm. A short description of the DM algorithm, described in greater detail elsewhere \cite{diffMap}, is included for completeness.

\subsection{EMC intensity reconstruction}

The EMC algorithm builds a model of the 3D intensity from a large collection of non-oriented, shot-noise limited diffraction data. The orientational classification of the data is probabilistic, where the data are assigned probability distributions in the rotation group and these are systematically refined so as to maximize the likelihood of the intensity model. The EMC algorithm comprises three steps:

\begin{description}

\item{E-step:} Expand the grid intensities into the tomographic representation: $W[\mathbf{q}]\to W_{ij}$.

\item{M-step:} Update the tomographic intensities by expectation maximization: $W_{ij}\to W'_{ij}$.

\item{C-step:} Compress the tomographic model back into a grid model: $W'_{ij} \to W'[\mathbf{q}]$. 
\end{description}
We use pseudocode to describe these steps in the next sections. The notation matches the theoretical discussion in Section \ref{EMsection}. Spatial frequencies are denoted by $\mathbf{q}$ and $\mathbf{p}$, detector pixel indices by $i$, rotation group samples by $j$, and $k$ is always a data index.

\subsubsection{E-step: model expansion}\label{sssec:Estep}

In the E-step the grid model of the intensities is expanded into a redundant tomographic representation (model) to make the expectation maximization step (M) easier. Intensities $W_{ij}$ in the tomographic model are treated as independent variables by the M-step. 

Each element of the tomographic model is associated with a particular detector spatial frequency $\mathbf{q}_i$ and rotation matrix $\mathbf{R}_j$. The $M_\mathrm{pix}$ frequencies $\mathbf{q}_i$ refer to the detector (or particle) in an arbitrary reference orientation; the $M_\mathrm{rot}$ matrices $\mathbf{R}_j$ are rotations relative to the reference orientation. The construction of the $\mathbf{q}_i$ for a simple square array detector is given in Section \ref{Expparamsection}. Our rotation matrices are generated from pre-computed lists of quaternions that sample the rotation group at the desired resolution (see appendix \ref{apd:rot}). We use linear interpolation to extract intensity values at the rotated spatial frequencies \mbox{$\mathbf{q}'=\mathbf{R}_j \cdot \mathbf{q}_i$}.

\subsubsection{M-step: expectation maximization}\label{ssec:Mstep}

The probabilistic classification of the data, and then their aggregation into an improved tomographic model, is performed in the M-step. Central in this process is the computation of the conditional probabilities $P_{jk}$. These are based on the current intensity model, $W_{ij}$. When the diffraction data (photon counts) $K_{ik}$ are averaged with respect to these probabilities (equation (\ref{Mrule})), the result is a tomographic model $W'_{ij}$ with increased likelihood. 

\begin{algorithm}[t] 
\SetLine
\SetInd{0.5em}{0.5em}

\caption{E-step: model expansion}
\BlankLine
\KwIn{grid model $W[\mathbf{q}]$, reference tomograph spatial frequencies $\mathbf{q}_i$, rotation matrices $\mathbf{R}_j$}		
\KwOut{tomographic model $W_{ij}$}				

\SetKwFunction{interpolateFrom}{Interpolate}

\BlankLine

\For{$j\leftarrow 1$ \KwTo $M_\mathrm{rot}$}
{
\BlankLine
	
	\For{$i\leftarrow 1$ \KwTo $M_\mathrm{pix}$}
	{
	$\mathbf{q}' \leftarrow \mathbf{R}_j \cdot \mathbf{q}_i$ 
		
	$W_{ij} \leftarrow  \interpolateFrom(W[\mathbf{q}], \mathbf{q}' ) $
	}
\BlankLine

}

\BlankLine
\Return $W_{ij}$
\BlankLine
\end{algorithm}

The most time-intensive parts of the computation, and indeed of the whole reconstruction algorithm, are the nested loops over $k$, $j$, and $i$ that would imply an operation count that scales as $M_\mathrm{data} \times M_\mathrm{rot} \times M_\mathrm{pix}$. However, the innermost loop, on the pixel index $i$, can be greatly streamlined in both places where it occurs by skipping all the pixels that have zero photons. We use a sparse representation of the photon counts that reduces the time scaling to $M_\mathrm{data} \times M_\mathrm{rot} \times N$, since most non-zero counts will be a single photon and the average total photon number is $N$.

\begin{algorithm}[t!] 
\SetLine
\SetInd{0.5em}{0.5em}

\caption{M-step: expectation maximization}
 
\BlankLine
 
\KwIn{tomographic model $W_{ij}$, data $K_{ik}$, rotation group weights $w_j$}
\KwOut{updated model $W'_{ij}$, mutual information $I$} 

\SetKwFunction{CondProb}{CondProb}

\BlankLine
$I \leftarrow 0$

\For{$j \leftarrow 1$ \KwTo $M_\mathrm{rot}$}
{
	$S_j \leftarrow 0$
	
	\BlankLine
	\For{$i \leftarrow 1 $ \KwTo $M_\mathrm{pix}$}
	{
		$W'_{ij} \leftarrow 0$

	}
\BlankLine
}

\BlankLine

\For{$k \leftarrow 1 $ \KwTo $M_\mathrm{data}$}
{
	\BlankLine
	
	\For{$j \leftarrow 1$ \KwTo $M_\mathrm{rot}$}
	{
	
		$P_{jk}\leftarrow \CondProb(W_{ij},K_{ik})$
		
		\BlankLine
		
		\For{$i \leftarrow 1$ \KwTo $M_\mathrm{pix}$}
		{
		
		$W'_{ij} \leftarrow W'_{ij} + P_{jk} K_{ik}$
		}
		\BlankLine
		
		$S_j \leftarrow S_j + P_{jk}$
		
		$I \leftarrow I + P_{jk} \cdot \mathrm{Log} (P_{jk}/w_j)$
	
	}
	\BlankLine
}
\BlankLine

\For{$j \leftarrow 1$ \KwTo $M_\mathrm{rot}$}
{
\BlankLine
	\For{$i \leftarrow 1 $ \KwTo $M_\mathrm{pix}$}
	{
		$W'_{ij} \leftarrow W'_{ij}/S_j$

	}
\BlankLine
}

\BlankLine

$I \leftarrow I/ M_\mathrm{data}$

\BlankLine
\Return $W'_{ij}$, $I$
\BlankLine
\end{algorithm}

Two copies of the intensity model are held in memory at any time: the current model for conditional probabilities, and the updated model obtained by photon averaging. In the innermost loop computations of the conditional probabilities only the logarithm of the current model is used. The actual computation in this most time-intensive step involves incrementing the conditional probabilities $P_{jk}$ by $\log{W_{ij}}$ for the pixels $i$ that recorded photons in diffraction pattern $k$ (or a multiple of this if multiple photons were recorded). After the conditional probability for a particular $k$ is computed, the second time-intensive step is executed. In this the updated model $W'_{ij}$ is incremented by $P_{jk}$, again, for only the pixels $i$ where photons were recorded (or a multiple of this).

The pseudocode shows how directly the mutual information $I(K, \Omega)|_W$ is computed from the conditional probabilities $P_{jk}$. In Section \ref{ssec:rotsample} we show how this quantity provides a useful diagnostic for reconstructions in addition to having intrinsic value as a measure of information. There are a few places not shown in the pseudocode that require special attention in the implementation. As an example, it is important to check for over/underflow in the computation of the conditional probabilities when the logarithms of the not yet normalized probabilities are exponentiated.

\begin{algorithm}[t] 
\SetLine
\SetInd{0.5em}{0.5em}

\caption{conditional probability}
 
\SetKwFunction{CondProb}{CondProb}

\BlankLine

\KwIn{data index $k$, tomographic model $W_{ij}$, data $K_{ik}$, rotation group weights $w_j$}
\KwOut{$P_{jk}= \CondProb(W_{ij},K_{ik})$} 

\BlankLine
$S \leftarrow 0$
\BlankLine

\For{$j \leftarrow 1$ \KwTo $M_\mathrm{rot}$}
{
	$P_{jk} \leftarrow \mathrm{Log}(w_j)$
	
	\BlankLine
	
	\For{$i \leftarrow 1 $ \KwTo $M_\mathrm{pix}$}
	{
		$P_{jk} \leftarrow P_{jk}+K_{ik} \,\mathrm{Log}(W_{ij})-W_{ij}$

	}
	\BlankLine
	
	$P_{jk} \leftarrow \mathrm{Exp}(P_{jk})$
	
	$S \leftarrow S+P_{jk}$

}
\BlankLine

\For{$j \leftarrow 1$ \KwTo $M_\mathrm{rot}$}
{
	$P_{jk} \leftarrow P_{jk}/S$
}
\BlankLine
\Return $P_{jk}$
\BlankLine
\end{algorithm}

\subsubsection{C-step: model compression}

The C-step (\ref{compress}) is the reverse of the model expansion, or E-step, and both of these operations use far less time than the M-step that comes in between. Over the course of multiple EMC iterations, the combination of C-step followed by E-step has the effect of making the tomographic model of the intensity, $W'_{ij}$, consistent with a 3D model $W'[\mathbf{q}]$ defined on a grid. We use linear interpolation (as in the E-step) when collapsing the tomographically sampled intensities
onto the grid.

Because of noise, the averaging of the data $K_{ik}$ in the M-step produces a tomographic model that does not respect the Friedel symmetry $W'[\mathbf{q}]=W'[\mathbf{-q}]$ when compressed to the grid model. This symmetry is restored at the conclusion of the C-step by replacing $W'[\mathbf{q}]$ and $W'[\mathbf{-q}]$ with their average.

\begin{algorithm}[!t] 
\SetLine
\SetInd{0.5em}{0.5em}

\caption{C-step: model compression}
\BlankLine
\KwIn{tomographic model $W'_{ij}$, reference tomograph spatial frequencies $\mathbf{q}_i$, rotation matrices $\mathbf{R}_j$}		
\KwOut{grid model $W'[\mathbf{q}]$}				

\SetKwFunction{GridNeighbors}{GridNeighbors}
\SetKwFunction{InterpolationWeight}{InterpolationWeight}
\SetKwFunction{symmetrize}{FriedelSym}

\BlankLine

\ForEach{$\mathbf{q}$ }
	{
			$W'[\mathbf{q}] \leftarrow 0$	
			
			$S[\mathbf{q}] \leftarrow 0$
	}
\BlankLine

\For{$j\leftarrow 1$ \KwTo $M_\mathrm{rot}$}
{
\BlankLine
	
	\For{$i\leftarrow 1$ \KwTo $M_\mathrm{pix}$}
	{
		$\mathbf{q}' \leftarrow \mathbf{R}_{j} \cdot \mathbf{q}_{i}$ 
		
		$ G  \leftarrow \GridNeighbors(\mathbf{q}')$
			
		\BlankLine
		\ForEach{$\mathbf{p} \in  G $}
		{
		
			 $ f \leftarrow \InterpolationWeight(\mathbf{q'}-\mathbf{p})$	
			 
			 $ W'[\mathbf{p}] \leftarrow W'[\mathbf{p}] + f \, W'_{ij} $	
			 
			$ S[\mathbf{p}] \leftarrow  S[\mathbf{p}] + f$ 
		
		}
		
	}
\BlankLine

	}
\BlankLine

\ForEach{$\mathbf{q}$ }
	{
			$W'[\mathbf{q}] \leftarrow W'[\mathbf{q}]/S[\mathbf{q}]$	
	}
\BlankLine
$W'[\mathbf{q}]  \leftarrow \symmetrize(W'[\mathbf{q}])$ 

\BlankLine
\Return $W'[\mathbf{q}]$ 
\BlankLine

\end{algorithm}

\subsection{Phase retrieval}\label{ssec:phaseretrieval}

We use the difference-map algorithm \cite{diffMap} to reconstruct the phases associated with the Fourier magnitudes obtained by the EMC algorithm, as well as the Fourier magnitudes in the central missing data region ($q<q_\mathrm{min}$).

Our pseudocode for the difference-map emphasizes its generic character as a method for reconstructing models subject to two constraints. One constraint, provided by the EMC intensities, is the Fourier magnitude constraint. The difference map implements this constraint by the operation \texttt{FourierProj}. \texttt{FourierProj} takes an arbitrary, real-valued input contrast $C$ and projects to another contrast $F$ called the Fourier estimate. The action of \texttt{FourierProj} is most transparent on the Fourier transforms of the input and output contrasts. In the data region $q_\mathrm{min}<q<q_\mathrm{max}$, the output $F$ inherits the Fourier phases of the input $C$ with the Fourier magnitudes provided by the square roots of the EMC intensities. In the central missing data region, $q<q_\mathrm{min}$, both the Fourier phase and magnitude of the input are preserved in the output. Finally, for spatial frequencies above the data cutoff, $q>q_\mathrm{max}$, the Fourier amplitudes of $F$ are identically set to zero.

The other difference map constraint is implemented by the operation \texttt{SupportProj}. When acting on an arbitrary real-valued input contrast $C$, \texttt{SupportProj} outputs the support estimate $S$.  The output $S$ is obtained by zeroing the contrast in $C$ outside the support of the particle and all the negative contrast within the support. Since the phase reconstructions are performed on exactly the same size grids as the EMC intensity reconstruction, the radius of the spherical support region implemented by \texttt{SupportProj} is the same dimensionless radius $R$ (increased by a few grid units) that defines our binary contrast test particles and degraded resolution biomolecules (Section \ref{ssec:support}).

\subsubsection{Difference map phase reconstruction}

The difference-map reconstruction begins with a randomly generated initial real-valued contrast $X$ and is otherwise completely deterministic. As $X$ is updated by the operations \texttt{FourierProj} and \texttt{SupportProj}, the corresponding Fourier and support estimates, $F$ and $S$, are generated. In reconstruction problems that reach fixed points $X^\ast$, where the magnitude $\epsilon$ of the update $\Delta X$ vanishes, either $F$ or $S$ can be output as the solution since they are the same when $\epsilon$ vanishes. This is not the case even for our phase reconstructions with simulated data. There are multiple sources of error that make it impossible for both constraints to be satisfied simultaneously. The intensity truncation for $q>q_\mathrm{max}$, for example, introduces a small inconsistency even when the diffraction data are oriented perfectly by the EMC algorithm. Of the two alternatives to choose for the reconstructed contrast, we use the Fourier estimate $F$ for reasons that will be clear below, when we discuss the modulation transfer function (MTF) \cite{MTF}.

\begin{algorithm}[t] \label{Algorithm:DifMap}
\SetLine
\SetInd{0.5em}{0.5em}

\caption{difference-map algorithm for phase reconstruction}

\SetKwFunction{FourierProj}{FourierProj}
\SetKwFunction{SupportProj}{SupportProj}
\SetKwFunction{iter}{IterationCount}
\SetKwFunction{error}{ErrorMax}
\SetKwFunction{RandomContrast}{RandomRealContrast}
\SetKwFunction{Append}{Append}

\BlankLine
\KwIn{constraint projections \FourierProj and \SupportProj}
\KwOut{real-valued particle contrast $C$, error series $E$ }

\BlankLine

$C\leftarrow 0$

$M\leftarrow 0$

$X\leftarrow \RandomContrast $

\BlankLine

\For{$i \leftarrow 1$ \KwTo \iter}
{

	\BlankLine

	$S \leftarrow  \SupportProj \left( X \right)$
	
	$F \leftarrow \FourierProj \left( 2\, S -  X \right)$

	\BlankLine
	
	$\Delta X \leftarrow F-S$
	
	$X \leftarrow X + \Delta X$

	\BlankLine

	$\epsilon \leftarrow \|\Delta X\|$
	
	$E \leftarrow\Append(E,\epsilon)$

	\BlankLine

	\If{$\epsilon< \error$}
	{
	\BlankLine
	
	$C \leftarrow C + F$
	
	$M\leftarrow M+1$
	
	\BlankLine}

}

\BlankLine

$C \leftarrow C/M$

\BlankLine
\Return $C$, $E$
\BlankLine

\end{algorithm}

The behavior of the difference-map error metric $\epsilon=\|F-S\|$ typically has two regimes in phase reconstructions. In the early iterations $\epsilon$ decreases nearly monotonically, thereby improving the consistency between the Fourier and support estimates. In the second regime $\epsilon$ is relatively constant, with small amplitude fluctuations suggestive of a steady-state. Because $\epsilon=\|\Delta X\|$ also measures the magnitude of the update, the iterated contrast $X$ and the estimates $F$ and $S$ are also fluctuating in this regime. To produce reproducible results, we average the Fourier estimate $F$ in the steady-state and call this the result of the phase reconstruction.

Since our spherical support is consistent with either enantiomer of the particle (and the intensity data does not distinguish these either), a successful reconstruction will be the inversion of the true particle in about half of all attempts that start with a different initial $X$. In those cases we invert the reconstruction before making comparisons.

\subsubsection{Modulation transfer function}\label{sssec:MTF}

The Fourier estimates $F$ have, by construction, always the same Fourier magnitudes in the data region $q_\mathrm{min}<q<q_\mathrm{max}$ (provided by the EMC intensities). This means that the fluctuations of $F$ at these spatial frequencies are purely the result of phase fluctuations. By averaging the difference-map estimates $F$ (after the steady-state is established), we are performing the average
\begin{equation}\label{MTF}
\mathrm{MTF}(\mathbf{q}) = \langle \exp{(i \phi_\mathbf{q})} \rangle,
\end{equation}
where $\phi_\mathbf{q}$ is the Fourier phase at spatial frequency $\mathbf{q}$. Phases that are reconstructed well and fluctuate weakly give $\mathrm{MTF}\approx 1$, while strongly fluctuating phases lead to a small $\mathrm{MTF}$. Since the degree of fluctuation is correlated with the magnitude of $\mathbf{q}$, we additionally perform a spherical average of $\mathrm{MTF}(\mathbf{q})$ to define a modulation transfer function that concisely conveys the quality of the phase reconstruction as a function of the resolution $q=|\mathbf{q}|$. 

\section{Simulations}\label{Simulationssection}

This section explores the conditions necessary to reconstruct particles in numerical simulations. We are primarily interested in understanding the behavior of the reconstruction algorithm as a function of the dimensionless particle radius $R$. For any $R$, the feasibility and quality of reconstructions depends critically on three additional parameters:

\begin{enumerate}
\item{} Does the average number of photons per measurement, $N$, satisfy the criterion (\ref{criterion}) on the reduced information rate?
\item{} Do the $M_\mathrm{rot}$ discrete samples of the rotation group provide a sufficient approximation of the continuous group for particles of the given complexity?
\item{} Are the total number of measurements $M_\mathrm{data}$ sufficient to reconstruct the particle with acceptable signal-to-noise? 
\end{enumerate}
Although the parameters $N$ and $M_\mathrm{data}$ are determined by the experiment while $M_\mathrm{rot}$ is algorithmic, this distinction is artificial when we recognize that both the physical and computational components of the imaging process are subject to limited resources. We have studied the effects of these parameters systematically by reconstructing the binary contrast test particles described in Section \ref{Testparticlessection}. These particles resemble biomolecules at a resolution above the atomic scale and can be generated for any $R$. Our simulations culminate in a desktop computer reconstruction of the GroEL protein complex (Section \ref{Testparticlessection}) at a resolution corresponding to $R=8$.

\subsection{Data generation}\label{ssec:data}

All our simulations begin with the construction of the contrast of a particle at a specified dimensionless radius $R$ (Section \ref{Testparticlessection}). After embedding the contrast on a grid with the chosen oversampling (usually $\sigma=6$), the squared Fourier magnitudes are computed as a model of the intensities. All particles in our simulations thus have only a single discernable structure/conformation at the measured resolution. 

Simulated data --- tables of photon counts --- are generated by repeatedly Poisson-sampling the intensity model at a set of spatial frequency samples specified by our detector parameters. In each simulated measurement all the spatial frequencies are rotated by a random element of the 3D rotation group. We generate uniform rotation group samples by uniformly sampling points on the surface of the unit sphere in four dimensions (quaternions) and mapping these to orthogonal matrices (equation (\ref{3x3})). The rotation element used to produce each measurement is not recorded. Because the data is simulated with uniform rotation group samples, the group weights used by the reconstruction algorithm are the uniform sampling weights (\ref{eqn:rotweights}). Since the rotated spatial frequency samples will fall between the grid points of the intensity model, interpolation is used to define the mean of each Poisson-sampled photon count. Finally, by normalizing the intensity model as described in Section \ref{ssec:signalstrength}, our data have the property that the mean photon number per measurement is $N$.

The number of data used in the reconstructions ($M_\mathrm{data}$) can be very large, sometimes exceeding $10^6$ when $R$ is large and $N$ is small. By using sparse records of the photon counts (Section \ref{ssec:Memscaling}), however, the total storage required for the data is not much greater than the storage needed for a single 3D intensity model having the corresponding signal-to-noise. The reconstruction algorithm has, of course, no access to the 3D intensity model that was used to generate the data.

\subsection{Convergence with rotation group sampling}\label{ssec:rotsample}

We argued in Section \ref{Rotationsection} that an adequate sampling of the rotation group, for reconstructing particles of dimensionless radius $R$, is obtained when the rotation group sampling parameter $n$ (edge-subdivisions of the 600-cell) matches this value. Whereas the proportionality $n\propto R$ is clear, we present here some additional assessments that support the simple rule $n\approx R$. While larger $n$ are even better, the $n^3$ growth in the memory used by the algorithm motivates us to identify the smallest $n$ that achieves good results.

\begin{figure}[t!]
\centering
\includegraphics[width=3.3in]{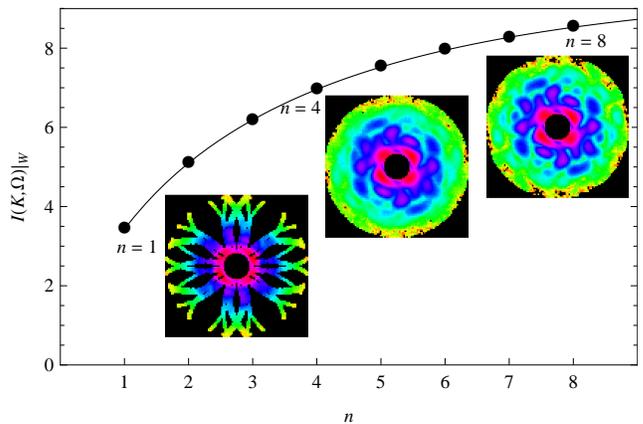} 
\caption{(Color online) Convergence with respect to rotation group sampling. The plot shows the increase in the mutual information $I(K,\Omega)|_W$ as the discrete sampling of the rotation group is increased; the integer $n$ is inversely proportional to the angular resolution. Saturation of the mutual information with $n$ indicates the data $K$ have exhausted the available orientational information in the intensity $W$, here for the case of an $R=8$ particle. The insets show corresponding cross sections of the intensity $W'$ generated by a single EMC update starting from the true intensity $W$. Speckles in the highest resolution shell appear at about $n=8$. The intensity scale is logarithmic and missing data regions are rendered black.  \label{fig:Ivsn}}
\end{figure}

The most direct test is to perform a single iteration of the EMC algorithm, beginning with the true intensity model. For this we generated data with sufficient total recorded photons ($N\times M_\mathrm{data}$) that signal-to-noise is not a factor. Since the data are generated by the same intensity model that begins the EMC update, the only thing that can spoil the preservation of the intensity by the update is the insufficient sampling of the rotation group. In Fig. \ref{fig:Ivsn} we show planar slices of the intensity model after one EMC update for a particle with $R=8$. The extreme case $n=1$, with only $M_\mathrm{rot}=60$ samples, is clearly inadequate because large regions of the intensity grid are never visited by a rotation of one of the detector's spatial frequency samples. This shortcoming is eliminated at about $n=4$ ($M_\mathrm{rot}=3240$), however, the intensity in the highest resolution shell lacks the expected speckle structure. These features first become established at level $n=8$ ($M_\mathrm{rot}=25680$).

There is another assessment of the rotation group sampling that does not require the true intensity model (or converged reconstruction). In this test we ask to what extent the data can detect additional rotational structure just by increasing the sampling parameter $n$. The measure of rotational structure most relevant to the available data is the mutual information $I(K,\Omega)|_W$. For a given intensity model $W$, this gives the information a typical measurement $K$ provides about its location in the rotation group. Clearly this depends on $W$ --- possibly a poor approximation of the true intensity --- as well as the noise in the data (mean photon number $N$). In any case, if the value of $I(K,\Omega)|_W$ is significantly increased upon increasing $n$, then there is information in the data that can detect additional rotational structure that should be exploited. In our implementation of the EMC algorithm $I(K,\Omega)|_W$ is calculated with negligible overhead in every iteration (see algorithm 4). To test convergence with respect to $n$ we simply increase $n$ and observe how much $I(K,\Omega)|_W$ increases. In Fig. \ref{fig:Ivsn} we also show the behavior of $I(K,\Omega)|_W$ as a function of $n$ for the same particle and data used to create the intensity cross sections. The leveling off at $n\approx 8$ is consistent with our earlier observations.

\subsection{Feasibility with respect to mean photon number}\label{ssec:N}

\begin{figure}[t!]
\centering
\includegraphics[width=3.3in]{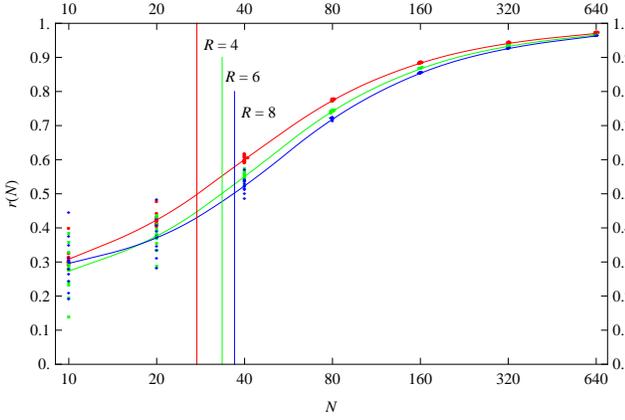} 
\caption{(Color online) Numerically computed reduced information rate $r(N)$ as a function of mean photon number $N$, for particle sizes $R=4,6,8$ (red, green and blue respectively). The interpolating curve shows mutual information averaged over 11 particles --- represented by the scattered points --- at each $R$ for various $N$. The vertical lines intersect the curves at their respective $r(N)=1/2$ points.\label{fig:criterion}}
\end{figure}

The total number of photons recorded in an imaging experiment is $N\times M_\mathrm{data}$. If the particle orientations were known for each of the diffraction data, then the quality of reconstructions would be independent of how the photon budget is allocated: simple signal averaging will give the same result if half the number of photons ($N/2$) are recorded on twice the number of data sets ($2 M_\mathrm{data}$). This changes when the orientations are unknown, and we rely on the reduced information rate $r(N)$ for guidance  (Section \ref{ssec:NoiseCriterion}).

We computed $r(N)$ using equation (\ref{criterion}) by numerically evaluating $I(K,\Omega)|_W$ for binary contrast test particles. The strict definition of $r(N)$ calls for an average over an intensity ensemble $W$; in our case this corresponds to particles of a particular radius $R$. Figure \ref{fig:criterion} shows plots of $r(N)$ as a function of the mean photon number $N$ for $R=4,6$, and $8$. As shown by the small scatter in the results for the larger $N$, fluctuations of $I(K,\Omega)|_W$ within each radius ensemble are very small, thus establishing $r(N)$ as a useful statistic when given only the particle radius. The single most important conclusion to draw from the $r(N)$ plots is that the reduced information rate is negligibly reduced (from unity) for even relatively small $N$. Taking $r(N)>1/2$ as the feasibility criterion, we obtain mean photon number thresholds of $N= 27.5, 33.5$, and $36.9$ for the three sizes of particles. From studies of the 1D minimal model \cite{Elser}, where this behavior can be analyzed in greater detail, we expect the threshold $N$ values to grow logarithmically with $R$.

\begin{figure}[t!]
\centering
\includegraphics[width=3.3in]{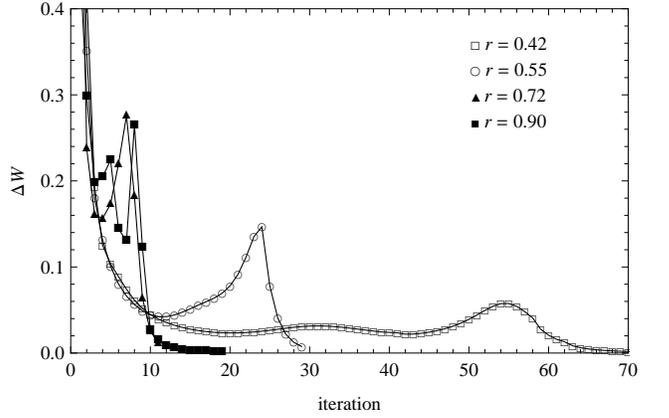} 
\caption{Update magnitudes $\Delta W$ for the intensity reconstruction of an $R=8$ particle at four values of the reduced information rate: $r(25) =  0.42$, $r(45) =  0.55$, $r(80) =0.72$ and $r(225) =0.90$. The normally rapid convergence ($\Delta W\to 0$) of the EMC algorithm becomes protracted as $r(N)$ approaches the value $1/2$. The corresponding particle reconstructions are shown in Fig. \ref{fig:res_vs_N}.  \label{fig:WvsN}}
\end{figure}

\begin{figure}[h!]
\centering
\includegraphics[width=3.3in]{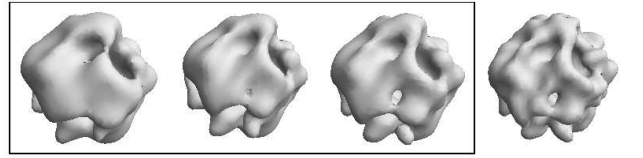} 
\vskip .5in
\includegraphics[width=3.3in]{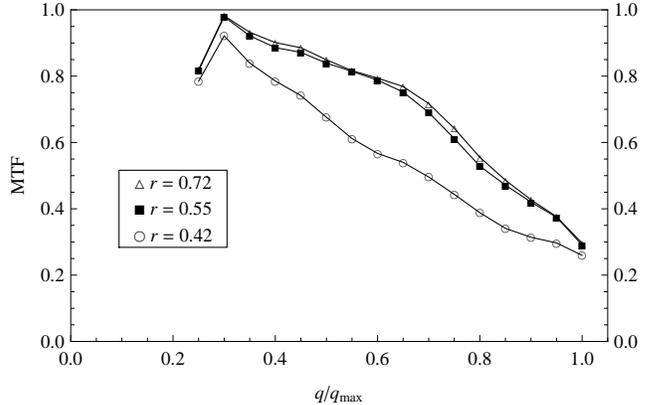}
\caption{Reconstructions of an $R=8$ test particle at three values of the reduced information rate $r(N)$, all other parameters unchanged. Shown in the top panel, from left to right, are $r(25) =  0.42$, $r(45) =0.55$ and $r(80) =0.72$; the true particle is reproduced on the right. The bottom panel shows the corresponding modulation transfer function (MTF) computed by the phasing algorithm. Behavior of the  EMC algorithm for these reconstructions is shown in Fig. \ref{fig:WvsN}. \label{fig:res_vs_N}}
\end{figure}

The consequences of $r(N)$ being below the feasibility threshold are noticed in two ways. First, there is a marked change in the behavior of the EMC intensity reconstruction algorithm, the progress of which we monitor by the time series of the update magnitudes:
\begin{equation}
\Delta W^2=\langle |W^\prime(\mathbf{p})-W(\mathbf{p})|^2\rangle_\mathbf{p}.
\end{equation}
Here $W^\prime$ is the update of $W$ and the average is uniform over all spatial frequencies between $q_\mathrm{min}$ and $q_\mathrm{max}$. Figure \ref{fig:WvsN} contrasts $\Delta W$ time series for four EMC intensity reconstructions of the same $R=8$ particle, the data differing only in the value of $N$ with all other parameters, including the total photon number, identical. The rapid decrease to zero, seen in the reconstructions with $r(80)=0.72$ and $r(225)=0.90$, is typical when $r(N)>1/2$. Likewise, the strongly non-monotonic behavior that stretches out over many iterations is normal for data that according to our $r(N)$ criterion is below the feasibility threshold. At the first broad minimum of $\Delta W$ in the plot for the case $r(45)=0.55$ the reconstructed intensity is nearly spherically symmetrical (a powder pattern). It takes the EMC algorithm many iterations to develop speckle structure by the gradual amplification of small features.

A second manifestation of being below our feasibility criterion is a loss of resolution in the final particle reconstruction. This is demonstrated in Fig. \ref{fig:res_vs_N}, which shows the results of three of the reconstructions described above. The degradation of high spatial frequency detail in the reconstructions at small $r(N)$ is a direct consequence of the reduced information rate in these data. With less total information available to reconstruct an accurate intensity, the resulting phase reconstruction of the particle is compromised.

\subsection{Reconstruction noise and number of measurements}\label{ssec:Mdata}

Even when the orientations of the diffraction patterns are known, shot noise in the intensity measurements will limit the signal-to-noise of the reconstructed particle. The resolution of the reconstruction will be compromised if the signal-to-noise at the highest spatial frequencies is poor. Because the intensities within a speckle are correlated, a natural quantity to consider is the total number of photons recorded in a typical, high spatial  frequency speckle. If we denote this number by $\mu$, the associated shot noise magnitude is $\sqrt{\mu}$, and the signal-to-noise in the highest frequency features of the reconstruction is $\sqrt{\mu}$. The scaling of $\mu$ with the experimental parameters is obtained by dividing the total number of recorded photons, $N \times M_\mathrm{data}$, by the number of speckles. Since the latter scales as $R^3$, the bound on the signal-to-noise given by oriented diffraction patterns and perfect phase reconstruction scales as
\begin{equation}\label{SNR}
\mathrm{SNR}\propto \sqrt{\frac{N M_\mathrm{data}}{R^3}}\propto \sqrt{\frac{N M_\mathrm{data}}{M_\mathrm{rot}}}=S,
\end{equation}
where the second proportionality follows from (\ref{eqn:Mrot}) and defines a convenient dimensionless measure of the signal. We see that $S^2$ is simply the average number of photons assigned, in a classification scheme, to each tomograph in a  tomographic representation of the 3D intensity when the rotation group is sampled at the appropriate resolution. 

\begin{figure}[t!]
\centering
\includegraphics[width=3.3in]{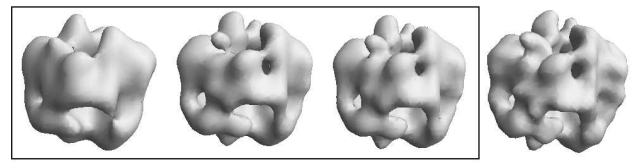}
\vskip .5in
\includegraphics[width=3.3in]{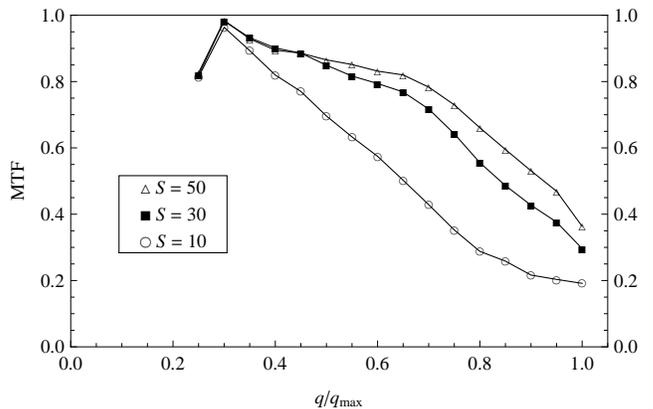} 
\caption{The same particle reconstructed with the same average number of photons per diffraction pattern, $N=100$, and increasing data. From left to right in the top panel are shown reconstructions with signal-to-noise parameter $S = 10,30,50$; the true particle is on the right. The panel below shows the corresponding modulation transfer function (MTF) computed by the phasing algorithm.
\label{fig:res_vs_S}}
\end{figure}

When the diffraction data are not oriented, the reduction in the information rate, as measured by $r(N)$, leads to a loss in signal-to-noise. For small data sets (which may not be sufficient to reconstruct the particle) this effect is modeled by replacing $M_\mathrm{data}$ with $r(N) M_\mathrm{data}$ in (\ref{SNR}). However, as argued in Section \ref{ssec:N}, for even modest $N$,  $r(N)$ is close to unity and loss of signal-to-noise by this mechanism is minor. 

Provided the photon numbers are reasonable, say $r(N)>1/2$, reconstructions from non-oriented diffraction data will fail for the same reason they fail for oriented data: the signal-to-noise is simply too small. This is shown in Fig. \ref{fig:res_vs_S}, where three $R=8$ test particles are reconstructed at the same $r(N)=0.75$ and decreasing values of $S$ (decreasing $M_\mathrm{data}$). We see that below $S\approx 30$ the resolution of the reconstructed particle is far less than the intrinsic resolution of the $R=8$ particle used to generate the data. The EMC algorithm succeeds at reconstructing a low resolution model of the particle at the smaller $S$ because the speckles at small spatial frequency may have sufficient numbers of photons when the speckles at high frequencies do not. We have defined $S$ so that the same standard, say $S\approx 30$, is meaningful for particles of arbitrary size.

\subsection{A biomolecule at 2 nm resolution}

We prepared an $R=8$ GroEL particle (1 nm half-period resolution) and simulated diffraction data from its Fourier intensities as described in Sections \ref{Testparticlessection} and \ref{ssec:data}, respectively. The mean photon number was set at $N=100$. For particles of size $R=8$ this implies a reduced information rate $r(N)=0.75$ (Fig. \ref{fig:criterion}). The EMC intensity reconstruction algorithm was run with rotation group sampling up to $n = 8$ ($M_\mathrm{rot}=25680$) and up to one million diffraction patterns ($M_\mathrm{data}=10^6$). All other parameters used in these simulations are listed in Sections \ref{Testparticlessection}, \ref{Expparamsection} and \ref{Reconparamsection}.

As discussed above (Section \ref{ssec:Mdata}), the EMC algorithm automatically reconstructs a lower resolution model when the number of data is such that only the speckles at low spatial frequency have adequate signal-to-noise. Because such lower resolution models have coarser angular features, a lower resolution sampling of the rotation group can be used to reconstruct them. These observations suggest a simple protocol for accelerating the solution process: first obtain a low resolution model and then refine this by increasing $M_\mathrm{rot}$ and $M_\mathrm{data}$ until the conditions for reconstructing up to the intrinsic resolution are reached. This strategy takes advantage of the fact that in practice very few EMC iterations are needed for the final, time-intensive, refinements. Time and memory scaling of the algorithm with $M_\mathrm{rot}$ and $M_\mathrm{data}$ are discussed in Section \ref{Scalingsection}.

\begin{figure}[t!]
\centering
\includegraphics[width=3.3in]{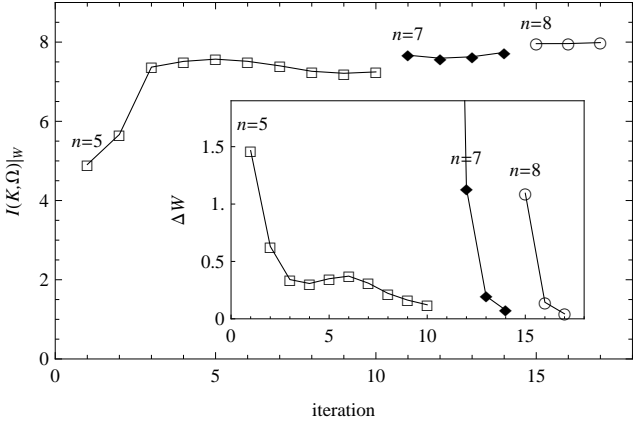} 
\caption{Mutual information $I(K,\Omega)|_{W}$ and EMC update magnitude $\Delta W$ as a function of iteration and increasing rotational sampling $n$ in the intensity reconstruction of the GroEL particle. The initial $n$ is small to save time and increased when the mutual information saturates and $\Delta W$ vanishes. With each increase of $n$ the number of data were also increased: $2\times10^5$, $5\times10^5$, and $10^6$. The intensity model after the 17th iteration was used in the phase reconstruction of the particle shown in Fig. \ref{fig:dens_GroEL}. \label{fig:cornELmI}}
\end{figure}

\begin{figure}[t!]
\centering
\includegraphics[width=3.3in]{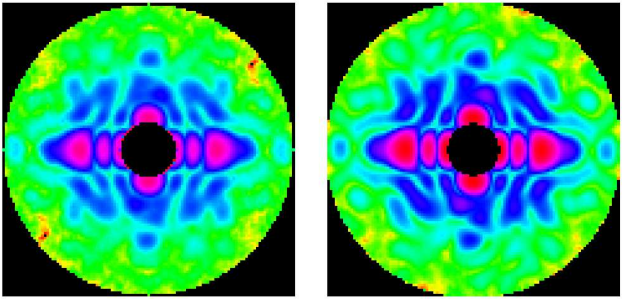} 
\caption{(Color online) Log-intensity cross sections after the $n=5$ (left) and $n=8$ (right) stages of the intensity reconstruction of the GroEL particle. The colors red (high) to yellow (low) span 4 orders of magnitude. The low frequency speckles are already fully reconstructed at rotational sampling $n=5$. \label{fig:intens_GroEL}}
\end{figure}

We began the intensity reconstruction of the GroEL particle using $n=5$ and $M_\mathrm{data}=2\times 10^5$ ($S=54$). Most of the time saving is the result of the reduced rotational sampling, which scales as $n^3$. A single EMC iteration on a 3 GHz machine at these parameter values takes 40 minutes. The reconstruction of low frequency speckles can be monitored by visually inspecting planar cross sections of the intensity model. A more quantitative approach makes use of the mutual information $I(K,\Omega)|_W$, where $W$ denotes the current intensity model. When this quantity saturates, the EMC algorithm is unable to improve the orientational accommodation of the data when evaluated at the limited rotational sampling. Stagnation of the EMC algorithm ($\Delta W =0$) also implies the likelihood function, of $W$ given the data, cannot be improved with the current settings. To improve both the orientational assignments of the data and the likelihood function, the rotational sampling and the number of data must be increased. Figure \ref{fig:cornELmI} shows the effects of gradually increasing $n$ and $M_\mathrm{data}$ on the value of $I(K,\Omega)|_W$. The refinements with $n=7$ and $n=8$ take very few iterations to reach the next (higher) plateau. We used the vanishing of $\Delta W$ as the stopping criterion for all the EMC stages, including the refinements. One EMC iteration in the final refinement stage ($n=8$, $S=62$) took about 24 hours. Figure \ref{fig:intens_GroEL} compares intensity cross sections after the early ($n=5$) and late ($n=8$) stages of rotational refinement.

\begin{figure}[t!]
\centering 
\includegraphics[width=3.3in]{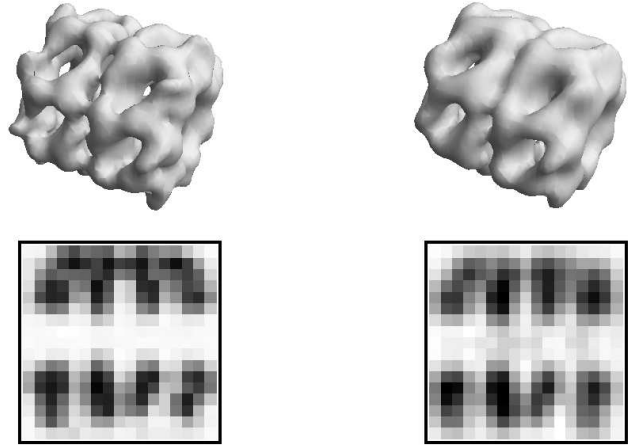} 
\caption{The $R=8$ GroEL particle (top left) compared with the results of our reconstruction (top right). The reconstruction was rotated to bring the two particles into alignment. The resolution of the reconstruction is degraded by about a factor of two relative to the model used to generate the data. Cross sections of the contrast are compared in the bottom row (left: model, right: reconstruction). \label{fig:dens_GroEL}}
\end{figure}

\begin{figure}[t!]
\centering
\includegraphics[width=3.3in]{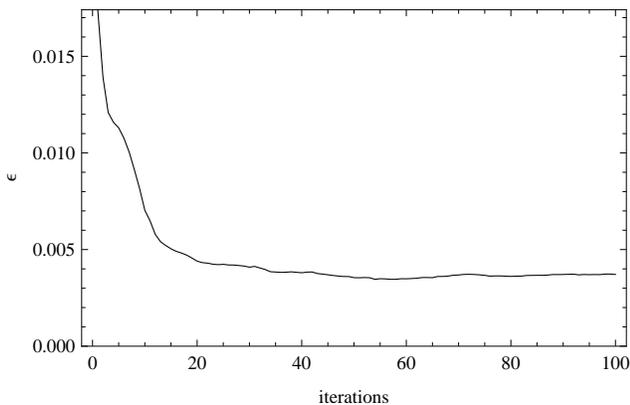} 
\caption{Difference map error metric $\epsilon$ in the phase retrieval of the GroEL intensity model obtained by the EMC algorithm. A steady state of residual phase fluctuations is reached after about 50 iterations. Averaging the phase reconstruction over the subsequent 200 iterations produced the particle contrast shown in Fig. \ref{fig:dens_GroEL} and the MTF function in Fig. \ref{fig:mtf_GroEL}.  \label{fig:err_GroEL}}
\end{figure}

The GroEL particle was reconstructed from the final EMC intensity model using the phase retrieval algorithm described in Section \ref{ssec:phaseretrieval}. The only other input to this algorithm is the support of the particle contrast and the constraint that the contrast is non-negative. Altogether there are four sources of error that can degrade the quality of the reconstructed particle. Three of these are responsible for errors in the intensity model: finite sampling of the rotation group, finite data sets, and grid interpolation errors (finite oversampling).  The fourth error source is the truncation of the data at spatial frequencies outside the range $q_\mathrm{min}$ to $q_\mathrm{max}$. The missing information for $q>q_\mathrm{max}$ has the greater effect, since the non-negativity constraint is very effective at reconstructing the missing beam-stop ($q<q_\mathrm{min}$) intensities when this region includes few speckles. Because the weak intensities beyond $q_\mathrm{max}$ are not reconstructed --- the algorithm treats them as zero --- a small incompatibility is introduced in the constraints used by the difference-map phasing algorithm. This together with the other sources of error lead to the nonvanishing of the difference-map error metric $\epsilon$ shown in Fig. \ref{fig:err_GroEL}, and averaging with respect to residual phase fluctuations is required to arrive at a reproducible result. The phase retrieval MTF function (Section \ref{sssec:MTF}) computed during the averaging period provides a comprehensive assessment of the internal self-consistency of the entire reconstruction process.

\begin{figure}[t!]
\centering
\includegraphics[width=3.3in]{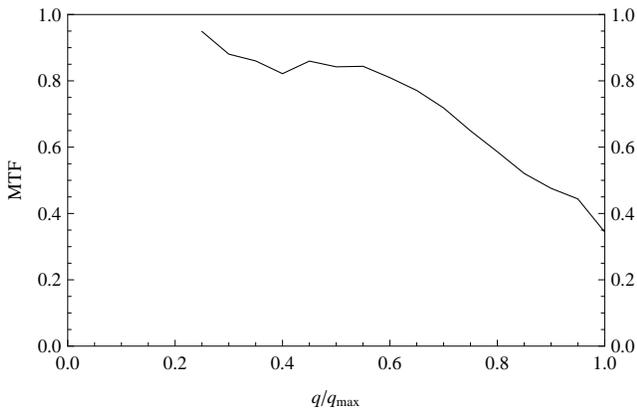} 
\caption{Phase retrieval MTF function for the GroEL particle reconstruction. \label{fig:mtf_GroEL}}
\end{figure}

The reconstructed $0.8$ MDa GroEL particle is compared in Fig. \ref{fig:dens_GroEL} with the $R=8$ resolution model used to generate the diffraction data. There is clearly a loss in resolution as a result of all the factors described above. From the phase retrieval MTF function, shown in Fig. \ref{fig:mtf_GroEL}, we see that contrast begins to deteriorate beginning with spatial frequencies about half the maximum of those measured (1 nm). Since the MTF begins to decline at about $0.5 \,q_{\text{max}}$, the reconstructed resolution is conservatively half the half-period resolution, or 2 nm. 

\section{Computational requirements}\label{Scalingsection}

Our simulations show that particles can be reconstructed at low resolution, $R<8$, and modest computational resources even when as few as $N=100$ photons are recorded on the average diffraction pattern. Because the parameters $R$ and $N$ are dictated by the physical properties --- including damage mechanisms --- of the sample and available light source and are therefore least under the control of the imaging experiment, it makes sense to assess the feasibility of real reconstructions as a function of these parameters. We will see that the computational resources are essentially independent of $N$ and scale as simple powers of $R$. This analysis assumes a fixed oversampling $\sigma$ and fixed signal-to-noise in the reconstructed contrast.

\subsection{Memory scaling}\label{ssec:Memscaling}

The data storage demands are modest and minor relative to the memory used by the algorithm when photon counts are recorded in a sparse format. A sparse-encoded measurement comprises on the order of $N$ integers identifying the detector pixels that have non-zero counts. For the pixels with single counts the pixel index provides a complete record; the small minority of pixels with multiple counts require an additional integer. Sparse data storage therefore scales in proportion to the total number of measured photons, $M_\mathrm{data}\times N$. At fixed signal-to-noise the number of photons aggregated per grid point in the intensity reconstruction is fixed. Since the number of points in the intensity grid scales as $R^3$, the total number of detected photons --- and the sparse storage space --- also scales as $R^3$. In a parallel implementation of the reconstruction algorithm it makes sense for all the data to be resident in each processor.

The largest set of variables used by the algorithm are the current and updated tomographic representations of the intensity models, $W_{i j}$ and $W_{i j}^\prime$. These have size $M_\mathrm{pix}\times M_\mathrm{rot}$ and scale as $R^2\times R^3=R^5$. The actual memory used depends strongly also on $\sigma$, which we assume is kept fixed as $R$ is varied. Our largest simulations ($R=8$, $\sigma=6$) used about 1 Gb with the memory dominated by these arrays.

\subsection{Time scaling}

The least certain part of our analysis is estimating the number of iterations needed by the EMC algorithm.  The simulations described in Section \ref{ssec:N} are consistent with the hypothesis that the number of iterations is fixed and small provided $N$ exceeds a modest information theoretic threshold (see Fig. \ref{fig:WvsN}). Because this minimum $N$ (see Fig. \ref{fig:criterion}) is believed to grow only logarithmically with $R$, we assume the criterion can always be satisfied and thus the number of EMC iterations is practically independent of $R$.

The most time-intensive operation in each iteration of the EMC algorithm is the expectation maximization step ($\mathbf{M}$) where the photon counts in each measurement, $K_{i k}$, are cross-correlated with the model log-intensities in each rotational sample, $\log{W_{i j}}$. The number of operations scales as $N\times M_\mathrm{data}\times M_\mathrm{rot}$ after the sum over the $i$-index, using the sparse data representation, is reduced to a sum of $N$ terms. As argued above, $N\times M_\mathrm{data}$ scales as $R^3$ and $M_\mathrm{rot}$ as $R^3$, giving a time per iteration that scales as $R^6$. This represents the time scaling of the reconstruction, since the number of iterations is independent of $R$ and the other two steps of the EMC update (E and C) are much faster because they do not involve the data.

\subsection{Parallel implementation}

Since both the memory and time scaling is dominated by operations on the tomographic intensity models $W_{i j}$ and $W_{i j}^\prime$, in a parallel implementation these and the cross-correlations on them would be distributed among the processors so as to minimize message passing. A natural approach is to have a separate master node perform the $\mathbf{E}$ step in such a way that blocks of the tomographic models (ranges of the $j$-index) are sent to different processors. After cross-correlating with all the data, which requires no message passing, each processor will send its results back to the master node for aggregation in the $\mathbf{C}$ step. In this scheme both memory per processor and time of the reconstruction are reduced in proportion to the number of processors. 

\section{Conclusions}

The aim of this study was to provide a detailed assessment of the feasibility and quality of reconstructions for the proposed single-particle imaging experiments by testing the performance of a particular algorithm developed for this purpose. The many dozens of reconstructions required to map out the parameters space could not have been carried out had the operation of the algorithm not become a fairly routine process. We never encountered a situation where the intensity reconstruction (EMC algorithm) had to be abandoned or restarted, or where the subsequent phase reconstruction did not reproduce the true particle to the expected resolution. Our attention to experimental details (\textit{e.g.} missing central data) in the simulations gives us confidence that the algorithm developed here will also succeed with real data. 

A variant of our algorithm was previously studied in the context of a minimal model having a single rotation angle as missing data \cite{Elser}. In the introduction we argued that in some respects the minimal model reconstructions might be harder than the reconstructions in a realistic model, where the diffraction data span the entire 3D rotation group. This scenario has been confirmed by our simulations. Recall that in the minimal model a 2D intensity distribution is sampled by non-intersecting 1D diffraction patterns, while in single-particle imaging the intensity is 3D and the 2D diffraction patterns intersect pairwise along arcs. In the minimal model the tomographic representation of the intensity is non-redundant, while in the 3D problem the tomographs are highly redundant and mutual consistency has to be imposed with the E and C steps of the EMC algorithm. The structure of the intersecting diffraction patterns and redundant variables in the 3D problem provide a mechanism, absent in the minimal model, that accelerates the reconstruction. The redundancy is greatest near the origin, where many diffraction patterns pass through the same speckles. Because the signal-to-noise is also greatest in this region, the reconstruction can begin there in a consistent fashion and then progress to higher resolution shells. By contrast, the rotational classification of diffraction patterns in the minimal model is not incremental and requires many more model-update iterations.

Reconstruction algorithms for the first round of experiments will have to address two additional complications not considered in our simulations. The first is the large shot-to-shot fluctuation in the incident photon flux of the source when the FEL process is unseeded \cite{non-seeded}. This adds another missing datum to the three orientational parameters, per diffraction pattern, that the expectation maximization step of the algorithm will have to reconstruct. The second complication is the background of photon counts arising from non-target particles, upstream beam optics and inelastic scattering. To deal with this the reconstruction algorithm should make use of the averaged dark count (no target particle). This represents additional data and requires a modification of the conditional probability computations.

\begin{acknowledgments}
This work was supported by Department of Energy grant DE-FG02-05ER46198 and was completed while the second author was a guest of the Centre for Free-Electron Laser Science at DESY, Hamburg. We would also like to thank the referees for their valuable patience and advice. 
\end{acknowledgments}

\appendix
\section{Mutual information identities}\label{apd:mi}

Given a trio of random variables $K$, $\Omega$ and $W$, we can evaluate the mutual information between one of them, say $K$, and the other pair, $(\Omega, W)$, treated as a single random variable. Writing the mutual information in terms of the entropy function $H$, we have
\begin{eqnarray}
\lefteqn{I(K,(\Omega, W)) = H(K)-H(K)|_{(\Omega, W)}}\\
&=&H(K)-H(K)|_\Omega+H(K)|_\Omega-H(K)|_{(\Omega, W)}  \nonumber\\
&=&I(K,\Omega)+I(K,W)|_\Omega
\end{eqnarray}
Interchanging $\Omega$ and $W$ in this derivation gives the identity
\begin{equation}
I(K,(\Omega, W))=I(K,W)+I(K,\Omega)|_W.
\end{equation}
Combining the two identities above we obtain the general result
\begin{equation}
I(K,\Omega)+I(K,W)|_\Omega=I(K,W)+I(K,\Omega)|_W.
\end{equation}
For our specific choice of random variables the mutual information $I(K,\Omega)$ vanishes identically because a measurement $K$ confers no information about the orientation $\Omega$ since the ensemble of models $W$ itself has an orientational degree of freedom that is uniformly distributed. Our identity thus involves only three terms:
\begin{equation}
I(K,W)|_\Omega=I(K,W)+I(K,\Omega)|_W.
\end{equation}

\section{Expectation maximization details}\label{apd:EM}

\subsection{Likelihood maximization}

Rearranging the order of the sums in the definition of the log-likelihood function we obtain
\begin{equation}\label{Qmax1}
Q(W^\prime)=\sum_{i=1}^{M_\mathrm{pix}}\sum_{j=1}^{M_\mathrm{rot}}(A_{ij} \log{W_{ij}^\prime}-B_j W_{ij}^\prime),
\end{equation}
where
\begin{eqnarray}
A_{ij} &=&\sum_{k=1}^{M_\mathrm{data}} P_{jk}(W) K_{ik}\\
B_j&=&\sum_{k=1}^{M_\mathrm{data}} P_{jk}(W).
\end{eqnarray}
Each term of the sum (\ref{Qmax1}) is of the form $a \log{W} - b\, W$ where $a$ and $b$ are positive constants. Since the terms are independent, the global maximum is achieved when each term is maximized with the value $W=a/b$.

\subsection{Fixed-point rotational invariance}

We wish to show that the intensities of the true model, $\widetilde{W}$, or more generally any rotation of this model, $\widetilde{W}^\mathbf{R}$, is a fixed point of the maximization update rule (\ref{Mrule}). Given the true model we can write down the probability distribution of the photon counts $K$:
\begin{equation}\label{P(K)}
P(K)=\sum_{j=1}^{M_\mathrm{rot}}w_j \,R_j(\widetilde{W}, K),
\end{equation}
where the joint Poisson distribution
\begin{equation}
R_j(\widetilde{W}, K)=\prod_{i=1}^{M_\mathrm{pix}} \frac{\widetilde{W}_{ij}^{K_{i}}}{K_i !}\exp{(-\widetilde{W}_{ij})}
\end{equation}
is the same (up to an irrelevant factor) as (\ref{Rjk}) but the data index $k$ has been replaced by the function argument $K$ representing an arbitrary vector of photon counts. Because the probability distribution on $K$ is unchanged if the model is rotated, we have
\begin{equation}\label{invariance}
\sum_{j=1}^{M_\mathrm{rot}}w_j \,R_j(\widetilde{W}^\mathbf{R}, K)=\sum_{j=1}^{M_\mathrm{rot}}w_j \,R_j(\widetilde{W}, K),
\end{equation}
for arbitrary rotations $\mathbf{R}$. The distribution (\ref{P(K)}) and the invariance (\ref{invariance}) are approximations that become exact in the limit $M_\mathrm{rot}\to\infty$.

Taking the numerator of the maximization update rule (\ref{Mrule}) and evaluating it for the rotated model $\widetilde{W}^\mathbf{R}$ with the data sum replaced by a sampling of $P(K)$, we have
\begin{eqnarray}
\lefteqn{\sum_{k=1}^{M_\mathrm{data}}P_{jk}(\widetilde{W}^\mathbf{R}) K_{ik}=}\\
& & M_\mathrm{data}\sum_K P(K) \left(\frac{w_j R_j(\widetilde{W}^\mathbf{R}, K)}{\sum_{j'=1}^{M_\mathrm{rot}}w_{j'} R_{j'}(\widetilde{W}^\mathbf{R}, K)}\right)K_i.\nonumber
\end{eqnarray}
Substituting (\ref{P(K)}) for $P(K)$ and using (\ref{invariance}), this reduces to
\begin{eqnarray}
\lefteqn{\sum_{k=1}^{M_\mathrm{data}}P_{jk}(\widetilde{W}^\mathbf{R}) K_{ik}}\nonumber\\
&=& M_\mathrm{data}\sum_K w_j \,R_j(\widetilde{W}^\mathbf{R}, K) \,K_i\\
&=& M_\mathrm{data}\, w_j \widetilde{W}^\mathbf{R}_{i j}
\end{eqnarray}
by the property that the mean of $K_i$ for the Poisson distribution is $\widetilde{W}^\mathbf{R}_{i j}$. By the same steps, the denominator of the update rule gives
\begin{equation}
\sum_{k=1}^{M_\mathrm{data}}P_{jk}(\widetilde{W}^\mathbf{R})=M_\mathrm{data}\, w_j,
\end{equation}
thus showing the desired fixed point property
\begin{equation}
\mathbf{M}:\quad \widetilde{W}^\mathbf{R}_{i j}\to \widetilde{W}^\mathbf{R}_{i j}.
\end{equation}

\subsection{Mutual information formula}

To evaluate the mutual information $I(K,\Omega)|_W$ using the quantities used by the EMC algorithm we need to approximate the integral over orientations $\Omega$ by a sum over the discrete samples $j$ and expectation values over photon counts $K_i$ by a normalized sum over the counts $K_{i k}$ in the actual data ($k$ is the data index).

Suppressing the model variables $W$, which we treat as a fixed quantity whenever $I(K,\Omega)|_W$ is calculated by the EMC algorithm, the mutual information is given by
\begin{equation}
I(K,\Omega)=\int_\Omega\, \sum_K\, P(K) P(\Omega | K) \log{\frac{P(\Omega | K)}{P(\Omega)}}.\label{miformula}
\end{equation}
Replacing the $\Omega$-integral by a weighted sum over samples $j$ and the $K$-expectation by a sum over the data, we obtain
\begin{equation}
I(K,\Omega)=
\sum_{j=1}^{M_\mathrm{rot}}w_j\, \frac{1}{M_\mathrm{data}}\sum_{k=1}^{M_\mathrm{data}}\,P(\Omega_j | K_k) \log{\frac{w_j P(\Omega_j | K_k)}{w_j}}.\nonumber
\end{equation}
We recover formula (\ref{eqn:MIKOmega}) with the identification
\begin{equation}
w_j P(\Omega_j | K_k)=P_{j k}(W).
\end{equation}

\section{Rotation group sampling based on the 600-cell}\label{apd:rot}

The quaternion parameterization of $3\times 3$ orthogonal matrices is given by the formula
\begin{equation}\label{3x3}
\mathbf{R}(q)=\left(
\begin{array}{ccc}
1-2 q_2^2-2 q_3^2&2q_1 q_2+2q_0 q_3&2q_1 q_3-2q_0 q_2\\
2q_2 q_1-2q_0 q_3&1-2 q_1^2-2 q_3^2&2q_2 q_3+2q_0 q_1\\
2q_3 q_1+2q_0 q_2&2q_3 q_2-2q_0 q_1&1-2 q_1^2-2 q_2^2
\end{array}
\right),
\end{equation}
where the unit quaternions $q=(q_0,q_1,q_2,q_3)$ are points on the unit 3-sphere. Because of the property $\mathbf{R}(q)=\mathbf{R}(-q)$, the unit quaternions are mapped 2-to-1 to the elements of the 3D rotation group. The multiplication rule for quaternions is most transparent in the $2\times 2$ spin-1/2 representation 
\begin{equation}\label{spinhalf}
q=q_0+i\,\mbox{\boldmath\( \sigma \)\unboldmath}\cdot \mathbf{q},
\end{equation}
($\mbox{\boldmath\( \sigma \)\unboldmath}$ are the Pauli matrices) which defines the ``scalar" ($q_0$) and ``vector" ($\mathbf{q}$) parts of the 4-vector. The scalar part encodes just the rotation angle, $\theta$, while the vector part also carries information about the rotation axis, $\mathbf{n}$:
\begin{equation}\label{angleaxis}
q_0=\cos{(\theta/2)}\qquad \mathbf{q}=\sin{(\theta/2)}\, \mathbf{n}.
\end{equation}
From (\ref{spinhalf}) and properties of the Pauli matrices one can verify that the inverse of a rotation is obtained by reversing the sign of the vector part. This fact is consistent with equations (\ref{3x3}) and  (\ref{angleaxis}).

Since the rotation required to move the element $q$ to the element $q^\prime$ is the quaternion $q^\prime q^{-1}$, a group-invariant distance between these elements should only be a function of the rotation angle (conjugacy class) of $q^\prime q^{-1}$, that is, its scalar part: $q_0 q_0^\prime+\mathbf{q}\cdot\mathbf{q}^\prime$. Since the latter equals
\begin{equation}
q\cdot q^\prime=1-\|q-q^\prime\|^2/2,
\end{equation}
the standard Euclidean distance on the quaternion 3-sphere is a group invariant distance between rotation group elements. This last remark is key for generating efficient samplings of the 3D rotation group: the unit quaternions should be placed to efficiently \textit{cover} the 3-sphere \cite{coverings}. A cover is optimal if it has the minimum number of points with the property that an arbitrary point of the space is always within a given covering radius $r_\mathrm{c}$ of some point in the cover. In our case the covering radius corresponds to a rotation angle $\delta\theta$, such that an arbitrary rotation group element is always within a rotation by $\delta\theta$ from one of our samples. If $q$ is a point of the cover and $q^\prime$ an arbitrary point, then
\begin{eqnarray*}
r_\mathrm{c}^2&>&\|q-q^\prime\|^2\\
&=&2-2q\cdot q^\prime\\
&=&2-2\left(q^\prime q^{-1}\right)_0\\
&=&2-2\cos{(\theta/2)}\\
&\approx&(\theta/2)^2.
\end{eqnarray*}
The covering radius, of the 3-sphere by quaternions, and the angular resolution $\delta\theta$ of the rotation group sampling are therefore related by:
\begin{equation}\label{rc}
r_\mathrm{c}\approx \delta\theta/2.
\end{equation}

Our covers of the 3-sphere are based on the highly symmetric 4-dimensional polytope that has the greatest number of regular tetrahedra --- 600 --- as its 3D facets. Known as the $\{3,3,5\}$ polytope, or 600-cell, it is the 4D analog of the regular icosahedron \cite{Coxeter}. The tetrahedral facets of the 600-cell are well covered by points arranged in the  fcc lattice. We will use the integer $n$ to describe the degree of the refinement of each tetrahedron by points of the fcc lattice. For example, when $n=4$, each tetrahedron edge is divided into 4 equal segments thus introducing $n-1=3$ edge points. This value of $n$ also introduces $(n-1)(n-2)/2=3$ points on each tetrahedron face and $(n-1)(n-2)(n-3)/6=1$ points in the interior of the tetrahedron. The $\{3,3,5\}$ polytope has 120 vertices, 720 edges, 1200 faces, and 600 cells. Combining this information with the point counts of the tetrahedron refinements, we obtain the following formula for the number of sample points on the 3-sphere:
\begin{eqnarray*}
\lefteqn{2M_\mathrm{rot}(n)} \\
&=& 120+720(n-1)+1200\frac{(n-1)(n-2)}{2} \\ 
& & +600\frac{(n-1)(n-2)(n-3)}{6}\\
&=&20(n+5n^3).
\end{eqnarray*}
The factor of 2 on the left side takes into account the overcounting by $\pm$ quaternion pairs.

To determine the appropriate level of refinement $n$, we need to compute the corresponding angular resolution $\delta\theta(n)$. Consider one tetrahedral cell of $\{3,3,5\}$; in canonical coordinates \cite{Coxeter} the vertices are
\begin{eqnarray*}
v_1&=&(1,0,0,0)\\
v_2&=&{\scriptstyle \frac{1}{2}}(\tau,1,1/\tau,0)\\
v_3&=&{\scriptstyle \frac{1}{2}}(\tau,1/\tau,0,1)\\
v_4&=&{\scriptstyle \frac{1}{2}}(\tau,0,1,1/\tau),
\end{eqnarray*}
where $\tau=(1+\sqrt{5})/2$ is the golden mean. The edge length of this tetrahedron is $1/\tau$; the tetrahedra of the refinement will have edge length $1/(n\, \tau)$ and this is also the minimum distance between fcc lattice points and $\sqrt{2}$ times the covering radius $r_\mathrm{c}$. When the fcc lattice points are projected to the unit 3-sphere by rescaling, the points near the center of the tetrahedron are expanded by the greatest amount. The linear expansion factor is given by the reciprocal of the distance between the origin and the tetrahedron center:
\begin{equation}
4/\|v_1+v_2+v_3+v_4\|=
\sqrt{8}/\tau^2
\approx1.080.
\end{equation}
The coarsest part of the sampling thus has minimum distance $r_\mathrm{c}=2/(n\,\tau^3)$. Using (\ref{rc}) we obtain:
\begin{equation}
\delta\theta(n)
=4/(n\,\tau^3)
\approx0.944/n.
\end{equation}

The samples of the 3D rotation group in this scheme carry non-uniform weights. Because the measure on the continuous group is just the volume element of the 3-sphere, the weight associated with a sample is proportional to the volume of its associated Voronoi cell. These weights/volumes are non-uniform as a result of two effects. First, there is a correction that affects the samples at the vertices and edges of the 600-cell, where the joining of the tetrahedral cells results in angular deficits. The second correction to the weights arises from the non-uniform distortion of the Voronoi cells, when these are projected from the 600-cell to the 3-sphere.

A regular (flat) tetrahedron has dihedral angle
\begin{equation}
\alpha=\cos^{-1}(1/3)\approx 70.5^\circ.
\end{equation}
Because five tetrahedra meet at every edge of the 600-cell, the fractional volume associated with samples on edges is given by
\begin{equation}
f_1=5\alpha/(2\pi)\approx 0.979566.
\end{equation}
The spherical angle subtended at a vertex of the regular tetrahedron, by spherical trigonometry, is $3\alpha-\pi$. Because 20 tetrahedra meet at every vertex of the 600-cell, each vertex sample has fractional volume
\begin{equation}
f_0= 20 (3 \alpha - \pi)/(4 \pi)\approx 0.877398.
\end{equation}
Since there are no deficits at samples on faces or within the cells of the polytope,
\begin{equation}
f_2=f_3=1.
\end{equation}

The volume change upon projecting a Voronoi cell from a $\{3,3,5\}$ facet to the unit \mbox{3-sphere} is the result of two things: (1) a uniform expansion by the linear scale factor $1/\|\tilde{q}\|$, where $\tilde{q}$ is the (non-unit) quaternion of the sample on $\{3,3,5\}$, and (2) projection of the 3-space of the tetrahedral cell to the tangent space of the 3-sphere at $q$, the projection of $\tilde{q}$. The second of these produces a reduction by the factor $q\cdot c$, where $c$ is the unit outward normal vector to the facet on which $\tilde{q}$ resides. In terms of the four cell-vertices,
\begin{equation}
c=\frac{v_1+v_2+v_3+v_4}{\|v_1+v_2+v_3+v_4\|}.
\end{equation}
The overall (unnormalized) weight of a sample $q$, originating from sample $\tilde{q}$ on the 600-cell, is given by
\begin{equation}
w(q)=f_k\, \frac{q\cdot c}{\|\tilde{q}\|^3} \label{eqn:rotweights},
\end{equation}
where $k=0,1,2,3$ is the associated dimensionality of the sample (vertex, edge, etc.). Vertex samples have the lowest weight, samples at the cell center the highest; their ratio is
\begin{equation}
\frac{w_\mathrm{vertex}}{w_\mathrm{center}}=\frac{f_0}{f_3}\left(\frac{\tau^2}{\sqrt{8}}\right)^4\approx 0.644 .
\end{equation}
In our pre-computed tables (for given $n$) of quaternion samples $q_j$ we include the weights $w_j\propto w(q_j)$ as the fifth component. All formulas in this paper assume the normalization
\begin{equation}
\sum_j w_j = 1.
\end{equation}

\bibliographystyle{unsrt}
\bibliography{spi3}

\end{document}